\documentclass[article]{aa}
\usepackage{graphicx}
\usepackage{txfonts}
\usepackage{natbib}
\usepackage{lscape}
\usepackage{multirow}

\bibpunct{(}{)}{;}{a}{}{,} % to follow the A&A style  
\newcommand{\myparagraph}[1]{\paragraph{#1}\mbox{}\\}
\begin{document}  

\title{Revisiting the cycle-rotation connection for late-type stars}  
  
\author{M. Mittag\inst{1} \and J.H.M.M. Schmitt\inst{1} \and K.-P. Schr\"oder\inst{2}}  
  
\institute{Hamburger Sternwarte, Universit\"at Hamburg, Gojenbergsweg 112,   
21029 Hamburg, Germany\\  
           \email{mmittag@hs.uni-hamburg.de}  
           \and   
           Departamento de Astronomia, Universidad de Guanajuato,  Callej\'{o}n de Jalisco s/n, 36023 Guanajuato, GTO, Mexico}
\date{Received \dots; accepted \dots}  
  
\abstract  
% Context  
{}   
% Aims  
{We analyse the relation between the activity cycle length
  and the Rossby number, which serves as a "normalised" rotation period 
  and appears to be the natural parameter in any cycle relation.}  
% Methods  
{We collected a sample of 44 main sequence stars with
  well-known activity cycle periods and rotation periods. To compute
  the Rossby numbers from the observed rotation periods, we used the
  respective B-V-dependent empirical turnover-times and derived
  the empirical relation between the cycle length and Rossby number.}  
% Results 
{We found a linear behaviour in the double-logarithmic relation
  between the Rossby number and cycle period. The bifurcation into
  a long and a short period branch is clearly real but it  depends, empirically,
  on the colour index $B-V$, indicating a physical dependence on effective
  temperature and position on the main sequence. Furthermore, there
  is also a correlation between cycle length and convective
  turnover time with the relative depth of the convection zone.
  Based on this, we derived empirical relations between cycle period
  and Rossby number individually for narrow $B-V$ ranges, for both
  cycle branches, as well as a global relation for the short-period branch.
  For the short period cycle branch relations, we estimated a scatter
  of the relative deviation between 14$\%$ and 28$\%$ on the long-period cycle
  branch. With these relations derived purely from stellar data, we obtained
  a good match with the 10.3$^{+1.1}_{-1.0}$ yr period for the
  well known 11-year solar Schwabe cycle and a long-period branch
  value of 104$^{+50}_{-34}$ yr for the 
  Gleissberg cycle of the Sun.
  Finally, we suggest that the cycles on the short-period branch
  appear to be generated in the deeper layers of the convective zone,
  while long-period branch cycles seem to be related to fewer deep layers in that zone.}  
% Conclusions  
{We show that for a broader $B-V$ range, the Rossby number is a more
  suitable parameter for  universal
  relation with cycle-rotation than just the
  rotation period alone. As proof, we demonstrate that our empirical stellar relations 
  are consistent with the 11-year solar Schwabe cycle, in contrast to
  earlier studies using just the rotation period in their relations.
  Previous studies have tried to explain the cycle position of the Sun in the
  cycle-rotation presentation via other kinds of dynamo, however, in
  our study, no evidence is found that would suggest another type of dynamo
  for the Sun and other stars.}

\titlerunning{Activity cycle estimation}  
  
\maketitle  
   
\section{Introduction}

The Sun is known to show activity cycles of very different lengths, the most
prominent ones being the so-called Schwabe and Gleissberg cycles.
The short-term cycle  discovered by \citet{schwabe1844}, has an average cycle
period of 11 years, and the other cycle,  first
reported by \citet{Gleissberg1939Obs}, has a  much longer timescale. 
Later, \citet{Gleissberg1945Obs}
estimated a cycle length of 77.7 yr for this solar longer-term cycle, but
other studies point to a length closer to 100 years (see below).  

In \citet{Richards2009PASP..121..797R}, the cycle lengths of the Schwabe
cycle are estimated from minima to minima and from maxima to maxima for
each cycle. A variation of the cycle length from 8.2 to 15 years
was found, with an average period of 11.0$\pm$1.5 yr
derived from the minima. Using the maxima,
the cycle length varies even more, from 7.3 to 17.1 years, but
the same average period of 11.0$\pm$2.0 yr was derived.  

In 1966, the systematic search for solar-like cycles in other stars
began with the Mount Wilson program \citep[see,][]{wilson1978}, monitoring 
the emission cores of the Ca~II H\&K lines. The so-called Mount Wilson
S-index (S$_{\rm{MWO}}$) measures the strength of the chromospheric
emission, which varies with the cycle period. The first S$_{\rm{MWO}}$ time
series were shown by \citet{wilson1978}, and from such time series, the
existence of solar-like activity cycles in other cool main sequence stars
was then demonstrated by \citet{Wilson1981SciAm}. 

 \citet{Noyes1984ApJ...287..769N} already used 13 such cycle periods
(found in Mount Wilson program stars) to study the cycle-rotation connection
and determined a empirical relation of $P_{\rm{cyc}} \propto P_{\rm{rot}}^{1.25}$.
This relation can also be interpreted in the form of
$P_{\rm{cyc}} \approx Ro^{1.25}$, where $Ro$
denotes the Rossby number, defined as the ratio of the rotation
period and the convective turnover time ($Ro=P_{\rm{rot}}/\tau_{\rm{c}}$),
as it may be assumed that the latter quantity is very similar for all
these 13 stars involved.

Using the cycle data presented by \citet{Noyes1984ApJ...287..769N},
 \citet{Baliunas1985ARA&A}, and \citet{Tuominen1988} studied the
correlation between the ratio of cycle length over rotation period
($P_{\rm{cyc}}/P_{\rm{rot}}$) and the fractional depth of the convection
zone, $d_{conv}$, and derived (theoretically) the
following relation: $P_{\rm{cyc}}/P_{\rm{rot}} \propto d_{conv}^{-0.5}$.
This same relationship was investigated empirically
by \citet{Saar1992}, who used a larger and updated cycle sample,
again obtained from the Mount Wilson time series. However,
a very much different dependence on the fractional depth, $d_{conv}$, was obtained
viz. $P_{\rm{cyc}}/P_{\rm{rot}} \propto d_{conv}^{-3.4}$, with an
exponent quite different from the theoretically derived 
value of \citep{Tuominen1988} (-0.5). Furthermore, \citet{Saar1992}
found a splitting into two cycle branches in the diagram of cycle period
versus dynamo number (which is the inverse quadratic Rossby number):
the older and thus less active stars are located on one branch,
while the younger and more active stars are located on the second
branch; therefore the two branches are known as the 'inactive' and 'active'
branches.

These findings were confirmed by \citet{Brandenburg1998ApJ}, who used
21 activity cycles with a very low false alarm probability (below $10^{-5}$),
taken from \citet{b95}. Here, the dimensionless ratio between
rotation period and cycle period introduced by \citet{Tuominen1988}
was used to study the relation between $P_{\rm{rot}}/P_{\rm{cyc}}$ and the
inverse Rossby number, as well as the relation between, $P_{\rm{rot}}/P_{\rm{cyc}}$, 
and the stellar activity indicator, $R_{\rm{HK}}^{\prime}$, for both branches.
With a larger cycle sample, \citet{Brandenburg2017} re-analysed the
latter relation, and in contrast to \citet{Tuominen1988}  \citet{Saar1992}, and
\citet{Brandenburg2017}  found no dependence of $P_{\rm{rot}}/P_{\rm{cyc}}$
on the fractional depth of the convection zone. 

A different attempt was made by \citet{boehm-vitence2007ApJ} to study
the cycle-rotation relationship. In her diagram of the cycle period versus rotation period,
two cycle branches are, again, visible. Furthermore,
\citet{boehm-vitence2007ApJ} found similar slopes on the inactive and
active branches, giving rise to her suggestion that the two different
branches are the signature of the operation of two different dynamos, such
that the cycles located on the inactive branch are created in a deeper
layer than the cycles located on the active branch.

Somewhat confusingly, the 11-year Schwabe cycle of Sun appears to be located
right in between those two cycle branches, cf., the $P_{\rm{cyc}}$
versus $P_{\rm{rot}}$ diagram of \citet[][Fig. 1]{boehm-vitence2007ApJ},
as well as in the $P_{\rm{rot}}/P_{\rm{cyc}}$ versus  $R_{\rm{HK}}^{\prime}$
diagram of \citet[][Fig. 4]{Brandenburg2017}.
\citet{boehm-vitence2007ApJ} therefore speculated that the solar cycle
could be influenced by both dynamo types of the two cycle
branches found in the stellar samples.
Another explanation was provided by \citet{Metcalfe2016ApJ...826L...2M},
who suggested the Sun to be in a transitional evolutionary phase
with the consequence that the 11-year solar activity cycle is
caused by a transitional dynamo, as well as considering that the Sun is not the only star
with a transitional dynamo operating, with other candidates, includng: HD~128620 and HD~166620 
\citep{Metcalfe2016ApJ...826L...2M,Metcalfe2017SoPh,Metcalfe2022ApJ.933L.17M}.

In this paper, we re-analyse cycle and rotation periods from data mostly provided by \citet{Brandenburg2017} to revisit
this relationship.
However, while \citet{Brandenburg2017} used the proportionality of 
$\omega_{\rm{cyc}}/\Omega \propto R_{\rm{HK}}^{\prime}$ to estimate cycle periods, which
takes into account the \ion{Ca}{II} excess fluxes, 
we go back to the approach of \citet{Noyes1984ApJ...287..769N} and simply
compare the empirical activity cycle period with the rotation period.

In addition, as developed below, we regard the Rossby number
to be the natural ingredient in this relation. It can be understood as
a normalised version of the rotation period, allowing us to consider
the effects of different convective turnover times along the
cool part of the main sequence. 
Finally, we compare the resulting empirical relations of our work with
those of previous studies, especially \citet{Brandenburg2017}.
Notably, these two different approaches in our way of picturing the
relationship between rotation and magnetic cycle length make a fundamental
difference: the two cycle branches obtained from the stellar data alone
are now fully consistent with the two main solar cycles, meaning that
the Sun is not a special case after all but it is mingled in our diagram
with all the other solar-like stars.
This now consistent picture encouraged us to also revisit the mentioned
relation between the ratio of cycle over rotation period and the 
depth of the convective zone, where the cycles of each branch originate.

\section{Theoretical considerations of the cycle-rotation connection}
\label{theory}

To explain the basic physics of the solar activity cycle, \citet{Parker1955}
was the first to introduce a simple theoretical dynamo model, which we refer
to as the $\alpha-\Omega$ dynamo in the following. Based on this model, \citet{Parker1955}
demonstrated the existence of migratory dynamo waves, which
can account for the basic features of the observed solar cycle.
The $\Omega$-effect is caused by the differential rotation of a star, which
stretches the magnetic field lines inside the star in a longitudinal direction
and produces toroidal magnetic field.  Next,
throughout a stellar convection zone, magnetic buoyancy causes magnetic
flux tubes to rise upwards, forming loops, which finally penetrate the
photosphere and produce dark, bipolar sunspot groups, until they eventually decay.  
By the action of Coriolis forces, the rising magnetic field tubes
are twisted and build up new poloidal magnetic fields, an effect that is commonly referred
to as the $\alpha$-effect today (but noting that \citet{Parker1955} actually used the
letter $\Gamma$ to describe this effect). As the $\alpha$-effect is
creating the poloidal field required for the next sunspot cycle, albeit
with inverted magnetic polarities, these two effects together are capable
of producing the observed quasi-periodic or cyclic magnetic field phenomena which seem to be
typical for solar and stellar  activity.

\citet{Stix1981} and \citet{Noyes1984ApJ...287..769N}, as well as the  references
given therein, provide an overview of simple dynamo models. For the critical
dynamo number $D_{crit}$,  \citet{Stix1981} derived the expression:
\begin{eqnarray} \label{dynamo_crit1}
D_{crit}  & \sim &  \frac{\alpha \Delta \Omega  R_{\star}^3}{\eta^2},
\end{eqnarray}  
where $\alpha$ denotes the `strength' of the $\alpha$-effect,
$\Delta \Omega$ is the change in angular velocity over the
length scale of turbulence $l$, $R_{\star}$ is the stellar radius, and $\eta$
is the turbulent diffusivity. By scaling in the usual fashion through
$\alpha \sim \Omega l$, $\Delta \Omega \sim  \Omega$ and
$\eta \sim l^2/\tau_c$ with $\tau_c$ (denoting the convective turnover time),
we find:
\begin{eqnarray} 
\label{dynamo_crit2}
D_{crit}  & \sim &  \left(\Omega  \tau_c \right)^2 \times \left(\frac{R_{\star}}{l}\right)^3.
\end{eqnarray}  

The first factor (on the right-hand side) of Eq.~\ref{dynamo_crit2} is simply the
inverse squared Rossby number and the second is the relative depth of turbulence.
Thus, Eq.~\ref{dynamo_crit2} suggests that dynamo action
depends on both rotation as well as stellar structure.

For the resulting magnetic cycle frequency, \citet{Parker1955}
derived the relation:
\begin{eqnarray}
 \label{parker_cycle_freq}
 \omega_{\rm{mag\_cyc}}  & = & \left(\frac{| k H \alpha|}{2}\right)^{\frac{1}{2}} ,
\end{eqnarray}  
where $H$ denotes the shear and Parker's original denotation $\Gamma$ is
replaced by $\alpha$.

For spherical geometry with radial shear, \citet{Stix1976IAUS} showed that 
by setting $H\approx r_{\odot}\delta \omega/\delta r$  and $k=2/r_{\odot}$,
Eq.~\ref{parker_cycle_freq} is equivalent to the expression:
\begin{eqnarray} \label{stix_cycle_freq}
 \omega_{\rm{mag\_cyc}}  & = & |\alpha \Omega^{\prime}|^{\frac{1}{2}}, 
\end{eqnarray}
where $\Omega^{\prime}$ is the angular velocity gradient.
Since both $\alpha$ and $\Omega^{\prime}$ scale with $\Omega$, it is clear
that the cycle frequency is expected to be proportional to the rotation
frequency, with the same, thus, being true for the respective periods.

As pointed out by \citet{Noyes1984ApJ...287..769N}, these expressions are
only valid in kinematic (i.e. linear theory) and
\citet{Noyes1984ApJ...287..769N} provide some examples of magnetic
dynamos, where cycle and rotations periods are not, in fact, linearly related.

Using the scaling for $\alpha \sim \Omega l$ and approximating the
shear through
 \begin{eqnarray} 
\label{shear}
H = \frac{d (r \Omega)}{dr} \sim \Omega,
\end{eqnarray}
that is, assuming small gradients in angular velocity and using periods,
we can rewrite Eq.~\ref{stix_cycle_freq}  as:
\begin{eqnarray} 
\label{stix_cycle_freq_rewrite}
  P_{\rm{mag\_cyc}} = 2P_{\rm{cyc}} & \approx & \sqrt{\frac{R_{\star}}{l}}P_{\rm{rot}}.
\end{eqnarray}
Equation~\ref{stix_cycle_freq_rewrite} thus shows that in the framework of this
simple ansatz, the magnetic cycle period should be directly proportional to
the rotation period, modified by with a factor which depends on stellar
structure. This factor, $l/R_{\star}$, is the relative (fractional) depth of the
turbulence, and with the assumption, the relative depth of the
turbulence is similar the relative depth of the convective zone. This relative depth is varied
in cool stars along the main sequence and we must expect an empirical dependence on effective temperature or
colour-index $B-V$.  

\begin{sidewaystable*}
\caption{Sample of used object and results}  
\label{tab1}
\centering
\begin{center}  
\begin{small}  
\scalebox{0.8}{
\begin{tabular}{lcccccccccccccccccccccc}  
\hline  
\hline  
\noalign{\smallskip}  
Name & B-V & S$_{\rm{MWO}}$ & Ref & $\log \rm{R}_{\rm{HK}}^{+}$ & Teff & [Fe/H] & age & $d_{cz}$ & P$_{\rm{rot}}$ [d] & Ref & $\tau_{\rm{c}}$ [d] & Ro & P$_{\rm{cyc}}^{S}$ [yr] & Ref & Cal. P$_{\rm{cyc}}^{S}$ [yr] & $\Delta$P$_{\rm{cyc}}^{S}/$P$_{\rm{cyc}}^{S}$ &  P$_{\rm{cyc}}^{L}$ [yr] & Ref & Cal. P$_{\rm{cyc}}^{L}$ [yr] & $\Delta$P$_{\rm{cyc}}^{L}/$P$_{\rm{cyc}}^{L}$\\
\hline  
\noalign{\smallskip}  
Sun & 0.642 & 0.169 & E17 & -5.05 &  5777 &  0.00 &  4.6$^{a}$ &  0.292 & 25.4$\pm$1.0 & B17 &  33.94 &  0.748 & 11.0$\pm$2.0 & B17 & 10.3$^{+1.1}_{-1.0}$ & 0.06 & 88.0 & P21 & 103.5$^{+49.8}_{-33.6}$ & -0.18 \\ 
HD~1835 & 0.659 & 0.349 & B95 & -4.30 &  5720 &  0.20 &  0.5$^{a}$ &  0.285 & 7.8$\pm$0.6 & B17 &  37.18 &  0.210 &  &   &  &   & 9.1$\pm$0.3 & B17 & 5.0$^{+2.7}_{-1.8}$ & 0.45 \\ 
HD~3651 & 0.85 & 0.176 & B95 & -5.10 &  5211 &  0.13 &  7.1$^{a}$ &  0.324 & 44.0 & B17 &  61.18 &  0.719 & 13.8$\pm$0.4 & B17 & 11.7$^{+1.2}_{-1.1}$ & 0.15 &  &   &  &   \\ 
HD~4628 & 0.89 & 0.23 & B95 & -4.81 &  5120 & -0.27 &  5.3$^{a}$ &  0.306 & 38.5$\pm$2.1 & B17 &  65.19 &  0.591 & 8.6$\pm$0.1 & B17 & 9.9$^{+1.2}_{-1.0}$ & -0.15 &  &   &  &   \\ 
HD~10476 & 0.836 & 0.198 & B95 & -4.93 &  5244 & -0.04 &  4.9$^{a}$ &  0.315 & 35.2$\pm$1.6 & B17 &  59.83 &  0.588 & 9.6$\pm$0.1 & B17 & 9.2$^{+1.1}_{-1.0}$ & 0.04 &  &   &  &   \\ 
HD~10780 & 0.804 & 0.28 & B95 & -4.57 &  5321 &  0.03 &  2.3$^{a}$ &  0.301 & 22.14$\pm$0.55 & O18 &  56.87 &  0.389 & 7.53$\pm$0.16 & O18 & 5.6$^{+0.9}_{-0.8}$ & 0.25 &  &   &  &   \\ 
HD~16160 & 0.918 & 0.226 & B95 & -4.86 &  5060 & -0.12 &  7.5$^{a}$ &  0.309 & 48.0$\pm$4.7 & B17 &  68.16 &  0.704 & 13.2$\pm$0.2 & B17 & 12.4$^{+1.3}_{-1.2}$ & 0.06 &  &   &  &   \\ 
HD~16673 & 0.524 & 0.215 & B95 & -4.65 &  6183 & -0.01 &  0.8$^{a}$ &  0.196 & 5.7 & N84 &  18.02 &  0.316 & 0.847$\pm$0.006 & M19a & 0.9$^{+0.2}_{-0.1}$ & -0.07 &  &   &  &   \\ 
HD~17051 & 0.561 & 0.225 & BS18 & -4.61 &  6045 &  0.15 &  1.0$^{a}$ &  0.239 & 8.5$\pm$0.1 & B17 &  21.98 &  0.387 & 1.6 & B17 & 1.4$^{+0.2}_{-0.2}$ & 0.13 &  &   &  &   \\ 
HD~18256 & 0.471 & 0.185 & B95 & -4.86 &  6395 &  0.05 &  3.2$^{a}$ &  0.145 & 3.65$\pm$0.03 & O18 &  13.56 &  0.269 &  &   &  &   & 6.64$\pm$0.06 & O18 &  &   \\ 
HD~20630 & 0.681 & 0.366 & B95 & -4.28 &  5654 &  0.06 &  0.6$^{a}$ &  0.279 & 9.2$\pm$0.3 & B17 &  41.83 &  0.220 &  &   &  &   & 5.6$\pm$0.1 & B17 & 6.4$^{+3.4}_{-2.2}$ & -0.14 \\ 
HD~22049 & 0.881 & 0.496 & B95 & -4.25 &  5140 & -0.09 &  0.6$^{a}$ &  0.306 & 11.1$\pm$0.1 & B17 &  64.27 &  0.173 & 2.9$\pm$0.1 & B17 & 2.6$^{+0.7}_{-0.5}$ & 0.12 & 12.7$\pm$0.3 & B17 & 8.5$^{+3.1}_{-2.3}$ & 0.33 \\ 
HD~26965 & 0.82 & 0.206 & B95 & -4.87 &  5282 & -0.29 &  7.2$^{a}$ &  0.299 & 43.0 & B17 &  58.33 &  0.737 & 10.1$\pm$0.1 & B17 & 11.5$^{+1.2}_{-1.1}$ & -0.14 &  &   &  &   \\ 
HD~30495 & 0.632 & 0.297 & B95 & -4.40 &  5804 &  0.09 &  1.1$^{a}$ &  0.268 & 11.4$\pm$0.2 & B17 &  32.16 &  0.354 & 1.7$\pm$0.3 & B17 & 1.6$^{+0.3}_{-0.2}$ & 0.06 & 12.2$\pm$3.0 & B17 & 16.0$^{+8.1}_{-5.4}$ & -0.31 \\ 
HD~32147 & 1.049 & 0.286 & B95 & -4.85 &  4801 &  0.30 &  6.4$^{a}$ &  0.336 & 48.0 & B17 &  83.93 &  0.572 & 11.1$\pm$0.2 & B17 & 11.7$^{+1.4}_{-1.3}$ & -0.06 &  &   &  &   \\ 
HD~37394 & 0.84 & 0.453 & B95 & -4.27 &  5234 &  0.09 &  0.6$^{a}$ &  0.300 & 10.78$\pm$0.02 & M17a &  60.21 &  0.179 &  &   &  &   & 5.83$\pm$0.08 & O18 & 8.4$^{+3.0}_{-2.2}$ & -0.44 \\ 
HD~43587 & 0.61 & 0.156 & B95 & -5.23 &  5876 & -0.03 &  4.2$^{a}$ &  0.271 & 22.6$\pm$1.9 & F20 &  28.58 &  0.791 & 10.44$\pm$3.03 & F20 & 10.4$^{+1.1}_{-1.0}$ & 0.0 &  &   &  &   \\ 
HD~75332 & 0.549 & 0.279 & B95 & -4.40 &  6089 &  0.10 &  0.4$^{a}$ &  0.223 & 4.8 & M19a &  20.60 &  0.233 & 0.493$\pm$0.003 & M19a & 0.5$^{+0.1}_{-0.1}$ & -0.04 &  &   &  &   \\ 
HD~75732 & 0.869 & 0.176 & B22 & -5.12 &  5167 &  0.34 &  5.2$^{a}$ &  0.334 & 37.4$\pm$0.5 & M17a & 63.05 & 0.593 & 10.9 & B22 & 9.7$^{+1.1}_{-1.0}$ & 0.11 &  &   &  &   \\ 
HD~76151 & 0.661 & 0.246 & B95 & -4.57 &  5714 &  0.11 &  1.6$^{a}$ &  0.285 & 15.0 & B17 &  37.58 &  0.399 & 2.5$\pm$0.1 & B17 & 2.4$^{+0.4}_{-0.3}$ & 0.02 &  &   &  &   \\ 
HD~78366 & 0.585 & 0.248 & B95 & -4.52 &  5961 &  0.03 &  1.1$^{a}$ &  0.243 & 9.7$\pm$0.6 & B17 &  24.99 &  0.388 &  &   &  &   & 12.2$\pm$0.4 & B17 & 8.9$^{+3.6}_{-2.6}$ & 0.27 \\ 
HD~100180 & 0.57 & 0.165 & B95 & -5.09 &  6013 &  0.00 &  2.3$^{a}$ &  0.243 & 14.0 & B17 &  23.06 &  0.607 & 3.6$\pm$0.1 & B17 & 3.4$^{+0.5}_{-0.4}$ & 0.06 & 12.9$\pm$0.5 & B17 & 20.4$^{+7.9}_{-5.7}$ & -0.58 \\ 
HD~100563 & 0.48 & 0.202 & B95 & -4.72 &  6357 &  0.09 &  2.3$^{a}$ &  0.157 & 7.73$\pm$0.04 & M19a &  14.23 &  0.543 & 0.609$\pm$0.009 & M19a &  &   &  &   &  &   \\ 
HD~103095 & 0.754 & 0.188 & B95 & -4.93 &  5449 & -1.35 &  4.6$^{a}$ &  0.228 & 31.0 & B17 &  52.52 &  0.590 & 7.3$\pm$0.1 & B17 & (9.6$^{+1.2}_{-1.0}$) & -0.32 &  &   &  &   \\ 
HD~114710 & 0.572 & 0.201 & B95 & -4.75 &  6006 &  0.06 &  1.8$^{a}$ &  0.243 & 12.3$\pm$1.1 & B17 &  23.31 &  0.528 &  &   &  &   & 16.6$\pm$0.6 & B17 & 15.6$^{+6.1}_{-4.4}$ & 0.06 \\ 
HD~115404 & 0.926 & 0.535 & B95 & -4.25 &  5043 & -0.16 &  1.4$^{a}$ &  0.300 & 18.5$\pm$1.3 & B17 &  69.03 &  0.268 &  &   &  &   & 12.4$\pm$0.4 & B17 & 14.5$^{+4.7}_{-3.5}$ & -0.17 \\ 
HD~120136 & 0.508 & 0.191 & B95 & -4.81 &  6245 &  0.28 &  0.4$^{a}$ &  0.195 & 3.05$\pm$0.01 & M17b &  16.54 &  0.184 & 0.333$\pm$0.002 & M17b & 0.3$^{+0.1}_{-0.1}$ & 0.04 &  &   &  &   \\ 
HD~128620 & 0.71 & 0.162 & BS18 & -5.16 &  5570 &  0.22 &  2.9$^{a}$ &  0.295 & 22.5$\pm$5.9 & B17 &  48.88 &  0.460 &  &   &  &   & 19.2$\pm$0.7 & B17 & 19.8$^{+5.6}_{-4.4}$ & -0.03 \\ 
HD~128621 & 0.9 & 0.209 & BS18 & -4.92 &  5098 &  0.24 &  4.7$^{a}$ &  0.329 & 36.2$\pm$1.4 & B17 &  66.24 &  0.547 & 8.1$\pm$0.2 & B17 & 9.2$^{+1.1}_{-1.0}$ & -0.14 &  &   &  &   \\ 
HD~140538 & 0.684 & 0.228 & M19b & -4.66 &  5645 &  0.05 &  2.7$^{a}$ &  0.280 & 20.71$\pm$0.32 & M19b &  42.51 &  0.487 & 3.88$\pm$0.02 & M19b & 4.5$^{+0.6}_{-0.5}$ & -0.16 &  &   &  &   \\ 
HD~146233 & 0.652 & 0.171 & M16 & -5.03 &  5741 &  0.04 &  3.5$^{a}$ &  0.286 & 22.7$\pm$0.5 & B17 &  35.81 &  0.634 & 7.1 & B17 & 7.2$^{+0.8}_{-0.7}$ & -0.02 &  &   &  &   \\ 
HD~149661 & 0.827 & 0.339 & B95 & -4.44 &  5265 &  0.04 &  2.0$^{a}$ &  0.301 & 21.1$\pm$1.4 & B17 &  58.98 &  0.358 & 4.0$\pm$0.1 & B17 & 5.3$^{+0.9}_{-0.8}$ & -0.32 & 17.4$\pm$0.7 & B17 & 17.5$^{+5.2}_{-4.0}$ & -0.01 \\ 
HD~152391 & 0.749 & 0.393 & B95 & -4.28 &  5462 & -0.02 &  0.8$^{a}$ &  0.294 & 11.4$\pm$1.4 & B17 &  52.11 &  0.219 &  &   &  &   & 10.9$\pm$0.2 & B17 & 9.3$^{+3.2}_{-2.4}$ & 0.15 \\ 
HD~156026 & 1.144 & 0.77 & B95 & -4.37 &  4633 & -0.18 &  1.4$^{a}$ &  0.311 & 21.0 & B17 &  97.60 &  0.215 &  &   &  &   & 21.0$\pm$0.9 & B17 &  &   \\ 
HD~160346 & 0.959 & 0.3 & B95 & -4.66 &  4975 & -0.02 &  4.4$^{a}$ &  0.320 & 36.4$\pm$1.2 & B17 &  72.75 &  0.500 & 7.0$\pm$0.1 & B17 & 9.0$^{+1.2}_{-1.1}$ & -0.29 &  &   &  &   \\ 
HD~165341~A & 0.86 & 0.392 & B95 & -4.38 &  5188 &  0.07 &  1.7$^{a}$ &  0.305 & 19.9 & B17 &  62.16 &  0.320 & 5.1$\pm$0.1 & B17 & 4.9$^{+0.9}_{-0.7}$ & 0.04 & 15.5$\pm$0.0 & B17 & 16.2$^{+5.0}_{-3.8}$ & -0.04 \\ 
HD~166620 & 0.876 & 0.19 & B95 & -5.02 &  5151 & -0.18 &  6.4$^{a}$ &  0.306 & 42.4$\pm$3.7 & B17 &  63.76 &  0.665 & 15.8$\pm$0.3 & B17 & 11.1$^{+1.2}_{-1.1}$ & 0.3 &  &   &  &   \\ 
HD~185144 & 0.786 & 0.215 & B95 & -4.79 &  5366 & -0.22 &  3.5$^{a}$ &  0.299 & 27.7$\pm$0.77 & O18 &  55.26 &  0.501 & 6.66$\pm$0.05 & O18 & 7.3$^{+1.0}_{-0.8}$ & -0.09 &  &   &  &   \\ 
HD~190406 & 0.6 & 0.194 & B95 & -4.81 &  5910 &  0.05 &  1.9$^{a}$ &  0.252 & 13.9$\pm$1.5 & B17 &  27.09 &  0.513 & 2.6$\pm$0.1 & B17 & 2.6$^{+0.4}_{-0.3}$ & -0.02 & 16.9$\pm$0.8 & B17 & 16.0$^{+6.3}_{-4.5}$ & 0.06 \\ 
HD~201091 & 1.069 & 0.658 & B95 & -4.30 &  4764 & -0.16 &  3.7$^{a}$ &  0.314 & 35.4$\pm$9.2 & B17 &  86.64 &  0.409 & 7.3$\pm$0.1 & B17 & 8.3$^{+1.3}_{-1.1}$ & -0.14 &  &   &  &   \\ 
HD~201092 & 1.309 & 0.986 & B95 & -4.52 &  4366 & -0.15 &  3.4$^{a}$ &  0.324 & 37.8$\pm$7.4 & B17 &  126.86 &  0.298 & 11.7$\pm$0.4 & B17 &  &   &  &   &  &   \\ 
HD~219834~B & 0.92 & 0.204 & B95 & -4.97 &  5055 &  0.21 &  6.2$^{a}$ &  0.329 & 43.0 & B17 &  68.38 &  0.629 & 10.0$\pm$0.2 & B17 & 11.0$^{+1.3}_{-1.1}$ & -0.1 &  &   &  &   \\ 
KIC~8006161 & 0.84 & 0.194 & BS18 & -4.96 &  5234 &  0.29 &  3.6$^{a}$ &  0.324 & 29.8$\pm$3.1 & B17 &  60.21 &  0.495 & 7.4$\pm$1.2 & B17 & 7.7$^{+1.0}_{-0.9}$ & -0.04 &  &   &  &   \\ 
KIC~10644253 & 0.59 & 0.219 & S16 & -4.65 &  5943 &  0.12 &  1.3$^{a}$ &  0.258 & 10.9$\pm$0.9 & B17 &  25.67 &  0.425 & 1.5$\pm$0.1 & B17 & 1.8$^{+0.3}_{-0.2}$ & -0.19 &  &   &  &   \\ 
\hline  
%\tableline  
\end{tabular}
}
\tablefoot{Here, the used objects are listed with the corresponding $B-V$ values from
  \cite{HIPPARCOS1997ESA}, except for the Sun. The solar $B-V$-value is taken from \citet{strobel1996}.
  Furthermore, we are listed the both activity indicators Mount Wilson S-index (S$_{\rm{MWO}}$) with 
  the reference and the corresponding $\log \rm{R}_{\rm{HK}}^{+}$ \citep{mittag2013} for each object,
  the effective temperature calculated with 
  $B-V$ relation from \citet[][Eq. 14.17]{Gray2005G}, the computed age with from the age relation
  by \citet[][label a: Eq. 12-14 \& label b: Eq. 3]{Mamajek-Hillenbrand2008ApJ},
  except the Sun, the solar age is taken from \citet[][label c]{2010NatGe...3..637B}, 
  and the median metalicity [Fe/H] from the catalogue by \citet[][Version 2020-01-30]{Soubiran2016},
  $d_{cz}$ the relative depth of the convection zone based on the $R_{cz}/R_{\star}$
  from catalogue by \citet{vanSaders2012ApJ}, the rotational period (P$_{\rm{rot}}$) and
  with the corresponding reference, the convective turnover time ($\tau_{\rm{c}}$) calculated with
  the relation of \citet{Mittag2018} and the corresponding the Rossby number (Ro),
  the cycle on the short period cycle branch period (P$_{\rm{cyc}}^{S}$) with reference, calculated
  cycle period (Cal. P$_{\rm{cyc}}^{S}$), the ratio of the difference between calculated and
  measured short period cycle, the cycle period (P$_{\rm{cyc}}^{A}$) on the long period cycle
  branch and the corresponding reference, calculated cycle period (Cal. P$_{\rm{cyc}}^{L}$) and
  the ratio of the difference between calculated and measured long period cycle. The
  calculated cycle value for HD~103095 listed in parentheses is computed
  with Eq. \ref{lin_log_pcyc_log_ro} and the corresponding parameter given in Table \ref{tab2}.}
\tablebib{(B95)~\citet{b95}, (B22)~\citet{Baum2022AJ}, (B17)~\citet{Brandenburg2017}, (BS18)~\citet{BoroSaikia2018yCat},
  (E17)~\citet{Egeland2017ApJ}, (F20)~\cite{Ferreira2020A&A}, (M16)~\citet{Mittag2016},(M17a)~\citet{Mittag2017b}, (M17b)~\citet{Mittag2017},
  (M19a)~\citet{Mittag2019a}, (M19b)~\citet{Mittag2019b}, (N84)~\citet{Noyes1984ApJ...279..763N}, (O18)~\citet{Olspert2018},
  (P21)~\citet{Ptitsyna2021Ge&Ae..61S..48P},
  (S16)~\citet{Salabert2016}, (V14)~\citep{Vidotto2014MNRAS.441.2361V}}
\end{small}  
\end{center}  
%\end{table*}  
\end{sidewaystable*}

\section{Data sample}

For this study, we used only main sequence stars with reliable cycle periods,
a well-observed rotation period, and  $B-V$ colours in the range
of 0.44 $<$ B-V $<$ 1.6, which is the valid definition range for the
transformation formula from Mount Wilson S-index into flux excess
$\rm{R}_{\rm{HK}}^{+}$ \citep{mittag2013} for main sequence stars.
An additional selection criterion is the Rossby number: it should be lower
than unity because by definition, that is, the Rossby number calculated with the empirical
convective turnover time as derived by \citet{Mittag2018} cannot
exceed unity in active main sequence stars. 
The reason for using only main sequence stars is to avoid mixing of
possibly different dynamo types. In total, we collected data for 44 stars, listed in Table \ref{tab1}.

The majority of these data was taken from \citet{Brandenburg2017} because,
in our opinion, this paper contains the best list of stars with reliable
cycle periods. In \citet{Brandenburg2017}, the solar Schwabe and Gleissberg
cycles are also listed, with 11$\pm$2 yr for the Schwabe cycle and 80 yr for
the Gleissberg cycle. However, both the length and the origin of the Gleissberg cycle are not
entirely clear. For example, \citet{Beer2018MNRAS} reported a slightly
longer Gleissberg cycle with  a period of 87~years, and a similar period for the Gleissberg cycle with 88 years was
found by \citet{Ptitsyna2021Ge&Ae..61S..48P}. \citet{Ptitsyna2021Ge&Ae..61S..48P}
also showed that the Gleissberg cycle has a triple-peak structure with periods of 60, 88, and 140 years.
In both of these studies, it was assumed that the origin of the Gleissberg cycle is the solar dynamo.
On the other hand, for example, \citet{Cameron2019A&A} determined a similar period of around 90~years
for the Gleissberg cycle, yet these authors argue that
cycles longer than the Schwabe cycle are probably randomly generated and not caused by any kind of
solar-like dynamo.

\begin{figure}  
\centering  
\includegraphics[scale=0.45]{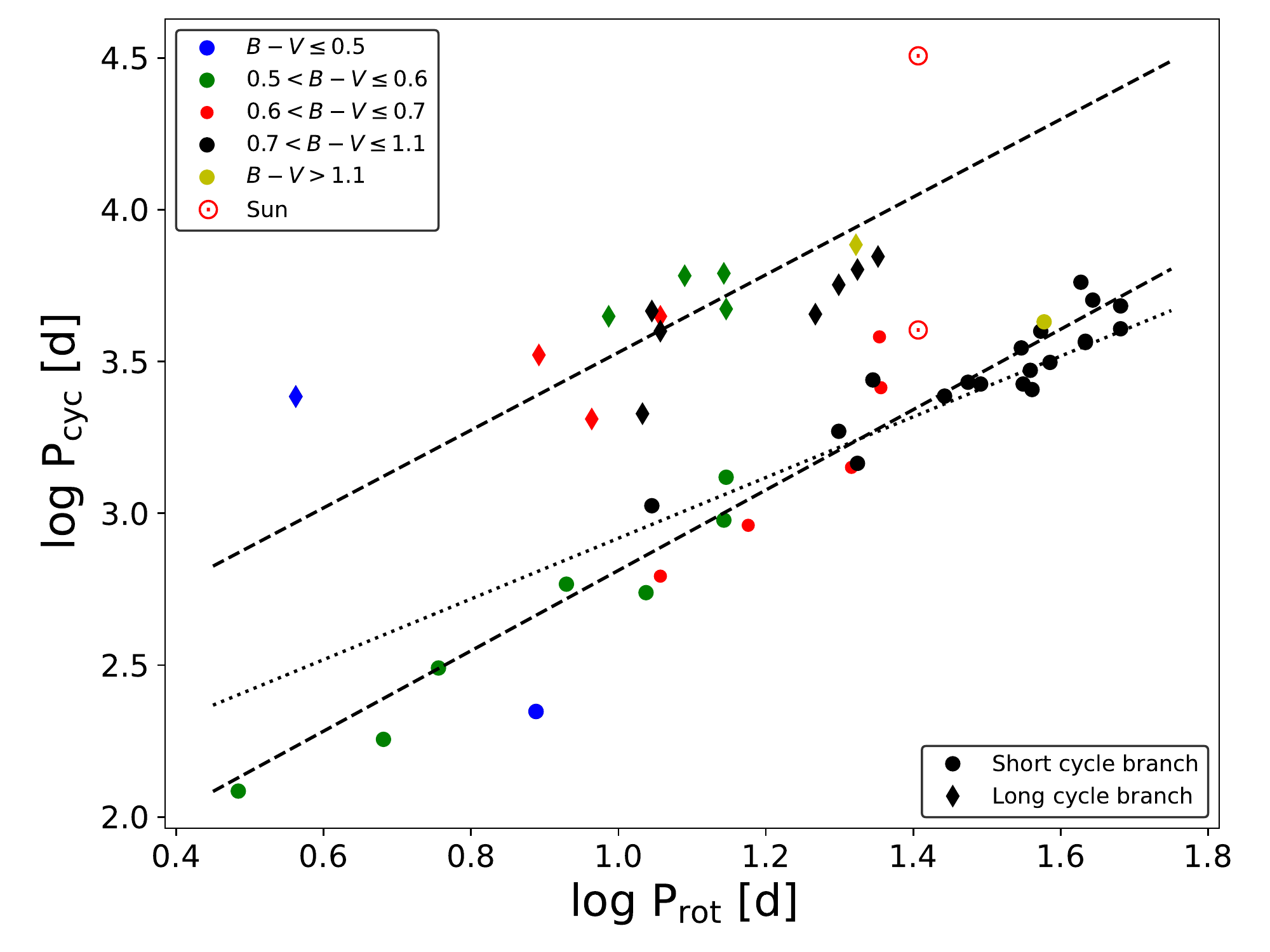}  
\caption{Logarithms of cycle period versus logarithms of rotation period for the sample stars.
  Here, the two branches are labelled with different signs and the different $B-V$ ranges are
  colour-coded. Furthermore, the dashed lines are depicted the estimated trends in the range
  $0.5 \le B-V \le 1.1$ of the cycle-rotation relations in both branches and the dotted line represents the
  trend with a fixed slope 1 for the short-cycle branch.}
\label{bv_plot}  
\end{figure}

In our study, we use the  Gleissberg cycle length of 88 years
found by \citet{Ptitsyna2021Ge&Ae..61S..48P} and assume
that the origin of the Gleissberg cycle is the solar dynamo.  Our assumption is motivated by the fact
that the period of the Gleissberg cycle is visible in different activity
indicators as in Sunspots and isotopes, furthermore, multi-cycle behaviour is also known
for other stars. In Fig. \ref{bv_plot}, we show the cycle periods versus the rotation periods in the
double logarithmic scale, which suggests that the Gleissberg cycle
could be part of the so-called 'active' branch or long-cycle branch, a term we  introduce later in this work.

We also remark that \citet{Brandenburg2017}  place the star HD~128620 (= $\alpha$ Cen A)
on their inactive branch (which we later denote as the short-cycle branch), although the assumed cycle
period of 19.2~years is actually located closer to the long-cycle branch. We decided to use the data as is
and assume the cycle period of 19.2~years to be a long-cycle period; we note in this context
that for HD~128620 the cycle periods have been derived from somewhat sparsely sampled
X-ray data, and not from Ca~II monitoring as is (typically) the case.

In addition, we include three stars (HD~10780, HD~37394, and HD~185144) from
the stars studied by \citet{Olspert2018}, which are not listed in
\citet{Brandenburg2017}.  \citet{Olspert2018} re-analysed the Mount
Wilson stellar sample and searched for activity cycles  using 
probabilistic methods. Here, we accept only those three stars where
the periods were confirmed by the BGLST method (Bayesian generalised
Lomb-Scargle periodogram with trend), as this method is similar
to the GLS method used by \citep{Zechmeister2009}.

HD~155886 basically fulfils those criteria as well. However, \citet{Olspert2018}
listed this star with two periods, one with five years and one with a 10.44 years,
with contradicting evidence which is the dominant one, and therefore
this star was not included in our sample.
We add three main sequence stars (HD~16673, HD~75332, and HD~120136)
to our sample, for which we have reported sub-yearly activity
cycles (see \citet{Mittag2017,Mittag2019a}) as well as the solar-like star
HD~140538 published by \citet{Mittag2019b}; these four F stars show
very cycle periods of high significance.

Finally, we added the stars HD~43587 and HD~75732 to our list. Both stars are also
listed by \citet{b95}, but no cycles were reported there.
The cycle period of HD~43587 was
published by \cite{Ferreira2020A&A}, and for HD~75732, the cycle period was published
by \citet{Baum2022AJ} from a study of  the combined S-index time series of a sample of 59 stars.
These stars were part of the Mount Wilson program and the Mount Wilson S-indices are
extended with S-values obtained with Keck observations obtained in the California Planet Search
program. \citet{Baum2022AJ} confirmed the cycle period reported by \citet{b95} with a cycle quality
classification of good or excellent; furthermore, \citet{Baum2022AJ} also published six cycle
periods not reported by \citet{b95}, but we have found only one measured rotation period
for HD~75732. Consequently, we did not consider the other cycle periods for our study of the
cycle-rotation period relation.
    
Most cycle periods of our sample stars fall on the `inactive' or
short-cycle branch in the cycle length over rotation period diagram,
which is clearly an observational bias. To find stellar analogues of
the Gleissberg cycle of the Sun, we need more stars with measured cycle
length of 50 years and above, which is impossible with the data currently available.
The stars HD~78366 and HD~114710 are notable exceptions, since these stars are listed with two cycle periods
in \citet{Brandenburg2017}, and both periods are located on the
`active', long-cycle branch. In \citet{Brandenburg2017}, the cycle
periods of these stars are briefly discussed, and from this discussion
we conclude that in both cases the shorter periods are uncertain, therefore,
we considered only the longer cycles in this study.

For all our sample stars, we derived the empirical convective turnover
time, $\tau_{\rm{c}}$, and the corresponding Rossby number, following
\citet{Mittag2018} (as mentioned above). Apart from these values, Table
\ref{tab1} also includes the effective temperature (Teff), based on the $B-V$
relation from \citet[][Eq. 14.17]{Gray2005G}, the age as expected from the
age relations by \citet[][Eq. 3 and Eqs. 12-14]{Mamajek-Hillenbrand2008ApJ},
and the median metalicity [Fe/H], using the values listed for the
corresponding star in the catalogue of
\citet[][version 2020-01-30]{Soubiran2016}.

In the framework of the general $\alpha-\Omega$ dynamo theory of
\citet{Stix1981}, the relative length-scale of turbulence ($l/R_{\star}$) plays
a role in the cycle-rotation relation. Assuming that the dynamo operates
at the base of the convection zone, the relative depth $l/R_{\star}$ is similar to
the relative depth of the convection zone. Thus, we can use
stellar structure models to obtain the latter value from 
the model grid for the convection zone  by \citet{vanSaders2012ApJ}.
We took the values of the base of the convection zones
($R_{cz}/R_{\star}$), using those with an initial model helium
abundance of 0.26 and the smallest difference of their effective temperature,
metalicity, and age values from the ones listed in Table \ref{tab1}.
From these base radii, we compute the respective relative depths of the
convection zones ($d_{cz}$=1-$R_{cz}/R_{\star}=l_{cz}/R_{\star}$), also listed in
Table \ref{tab1}.  

In our study, we used the denotation used by \citet{Brandenburg2017}, where
the 'inactive branch' was identified as a short-cycle branch and the
'active branch' as a long-cycle branch. Accordingly, we label the cycle periods 
located on the short-cycle branch as $P^{S}_{cyc}$ and those on the long-cycle 
branch as $P^{L}_{cyc}$ (see Sec. \ref{discussion} for further reasoning on 
this point).

We plot the cycle periods and the corresponding rotation
period in Fig.~\ref{bv_plot} using a double-logarithmic scale because the
Gleissberg cycle with 88~years is much longer than the other
cycles. In Fig.~\ref{bv_plot}, the two known cycle branches become clearly visible
and we use dots for the cycles on the short-cycle branch and diamonds for the cycles on the long-cycle branch
to distinguish both cycle branches.  We also colour-coded different $B-V$ ranges 
and this splitting up in the used  different $B-V$ ranges is the result of our
investigation (see Sect. \ref{per_rot_rossby_rel}).
Finally, we show the determined general trends of the cycle-rotation relation in both branches
estimated in the range of $0.5 \le B-V \le 1.1,$ 
as dashed lines in Fig.~\ref{bv_plot}; these trends are discussed in Sect. \ref{cyc_per} and
\ref{cyc_long_per}.

\section{The empirical cycle-rotation connection}
\label{cyc-rot_sec}
In this section, we study the connection between the observed activity
cycle and the rotation periods. We note that the number of available data
points are different for the aforementioned two branches. In total,
we have 34 cycles on the short-cycle branch, yet on the long-cycle branch,
we have only 17 cycles at our disposal. Accordingly, the relations for
the latter are far more uncertain. Nevertheless, for the sake of
completeness, we include also the long-cycle branch in this study.
Furthermore, for our study, we used only the data in the $B-V$
range of $0.5 < B-V \le 1.1$ for reasons described in Sect. \ref{per_rot_rossby_rel};
this range contains the vast majority of our data,
and inside this reduced $B-V$ range, we have 32 cycles 
on the short-cycle branch and 15 on the long-cycle branch. 

We begin this investigation with the short- and long-cycle branches and in trying to find
its empirical relation between the rotation and cycle periods. For this, we try a linear
relation between the logarithmic cycle period and logarithmic rotation periods.
Physically, this tests a power law relation between these
quantities as such. For the question, whether any additional parameters
would be needed, we then perform a principal component and a factor
analysis. Different approaches are compared as how the arrive at
statistically acceptable descriptions, by fixing the slope between cycle
and rotation periods at unity on the one hand and letting the slope vary
freely on the other hand, and introducing additional parameters. For these
studies, we use only the cycles on the short-cycle branch because the number of
cycles on the long-cycle branch is rather limited.

\subsection{Power law relations between $P_{cyc}$ and $P_{rot}$}

Here, we briefly discuss the power law relations between the cycle and
rotation periods in both cycle branches, which become linear relations in
the logarithmic variables.

\subsubsection{Short-cycle branch}
\label{cyc_per}

We begin our study with the empirical relation between the
(logarithmic) cycle lengths and rotation periods,
with the aim to check, whether $P^{S}_{cyc} \propto P_{rot}$ is valid, as suggested
by Eq.~\ref{stix_cycle_freq_rewrite} (with the index ${S}$ labelling
the periods on the short-cycle period branch).
\citet{Noyes1984ApJ...287..769N} used the inverse cycle period and
inverse Rossby number to study the cycle-rotation relation and derived
the relation $P^{S}_{\rm{cyc}} \propto P_{\rm{rot}}^{1.25\pm0.54}$, which is
consistent within the uncertainty with the theoretical prediction
by \citet{Stix1976IAUS} and also by \citet{Tuominen1988}.

In Fig.~\ref{bv_plot}, we plot (in a double-logarithmic representation)
the measured cycle periods versus the rotational periods 
of our sample stars. 
Since the theoretical expectation given by
Eq.~\ref{stix_cycle_freq_rewrite} suggests a linear relation viz:
\begin{eqnarray}
\label{lin_log_pcyc_log_ro_a}
  \log P_{\rm{cyc}} & \approx & a+n \log P_{\rm{rot}},
\end{eqnarray}
we then test whether the empirical slope is consistent with $n=1$.
Consequently, we fix $n$ at unity and optimised only the constant $a$ via
a leasts-square fit, obtaining a=1.918$\pm$0.027; the resulting relation
is illustrated by the dotted line in Fig.~\ref{bv_plot}.

Clearly, this regression with $n = 1$ does not fit the short periods. 
Next, we perform a least-squares fit treating also the slope $n$
as a free parameter, and obtain the relation
\begin{eqnarray}\label{pcyc-prot}
\log P^{S}_{\rm{cyc}}[d] = (1.488\pm0.092)+(1.324\pm0.067) \log P_{\rm{rot}}[d],
\end{eqnarray}
and plot this as the dashed line in Fig.~\ref{bv_plot}.
To quantify the remaining scatter of this best-fit empirical relation,
we compute the relative deviation
($(P^{\rm{S}}_{\rm{cyc}}-cal(P^{\rm{S}})_{\rm{cyc}})/P^{\rm{S}}_{\rm{cyc}}$ =
$\Delta$P$_{\rm{cyc}}^{S}/$P$_{\rm{cyc}}^{S}$),
where $cal(P^{\rm{S}})_{\rm{cyc}}$ is the calculated cycle period
and obtain a standard deviation ($\sigma$) of 0.24. 

Furthermore, we calculated the expected solar cycle period
with this relation and obtain a period of 6.1$^{+2.2}_{-1.6}$ yr, which
differs from the well-established real mean solar cycle period of 11~years.
Furthermore, the best-fit slope of 1.324 is obviously inconsistent
with theoretical considerations described in Sect.~\ref{theory}.

\subsubsection{Long-cycle branch}
\label{cyc_long_per}

We now turn to the relation between cycle periods and rotation periods
for the cycles located on the long-cycle branch.  In Fig. \ref{bv_plot}, the
available cycle and rotation period measurements are also plotted on a double-logarithmic scale.
As for the cycles on the short-cycle branch, we assume a linear relation between
the (logarithmic) quantities and using a least-squares fit, we arrive at the expression:
\begin{eqnarray}\label{pcyc_long-prot}
\log P^{L}_{\rm{cyc}}[d] = (2.22\pm0.38)+(1.28\pm0.33) \log P_{\rm{rot}}[d],
\end{eqnarray}
depicted as a dashed line in Fig.~\ref{bv_plot}. Next, we compute
the relative deviation ($\Delta P_{\rm{cyc}}^{L}/P_{\rm{cyc}}^{L}$) and obtain a standard deviation
of 40$\%$, a value about a factor of $\sim$2 larger than the standard deviation
on the short-cycle branch. Furthermore, the absolute
value of the slope is consistent with the slope in the relation for the short-cycle
branch (see Eq. \ref{pcyc-prot}).

Therefore, we conclude that the slope of main trends for 
both branches are equal and a similar result was found by \citet{boehm-vitence2007ApJ}.
However, the error of the slope is so large that this
slope is also consistent with unity within the error margin. To classify and assess
this large error, the small number of used cycle periods must be taken into account; thus,
it is important to find more cycles that include the corresponding rotation periods
located on the long-cycle branch.

Finally, we computed the expected solar Gleissberg cycle using our derived
relation (see Eq. \ref{pcyc_long-prot}) and obtained a cycle period
of 31$^{+91}_{-23}$ yr; the large error 
of this period reflects the large uncertainties of the cycle-rotation
relation for the long-cycle branch and, hence, the quantitative predictions are not particularly
meaningful.  
  
\subsection{Principal component and factor analysis of the
  $\log~P^{S}_{cyc}$ over $\log~P_{rot}$ relation}
\label{FA_PCA}

We now consider the question of whether additional parameters are required
to describe the relation between cycles and rotation periods.  From
Fig.~\ref{cyc_vs_prot}, it is obvious that different spectral types cover
different period ranges and,  furthermore,  the empirical relations 
between cycle length and rotation period taken in individual B-V ranges
differ from each other. This suggests the possibility of two kinds of 
effects of $B-V$: a dependence of the rotation periods themselves on $B-V$
and a dependence of the observed relation between cycle length and rotation
period. 

To test whether any such $B-V$ dependence is significant or not, we perform
a principal component and factor analysis. 
Factor analysis is a well established method to investigate how many
independent (or latent) variables exist in a given data set.  In our
specific case, we have at our
\begin{figure}  
\centering  
\includegraphics[scale=0.5]{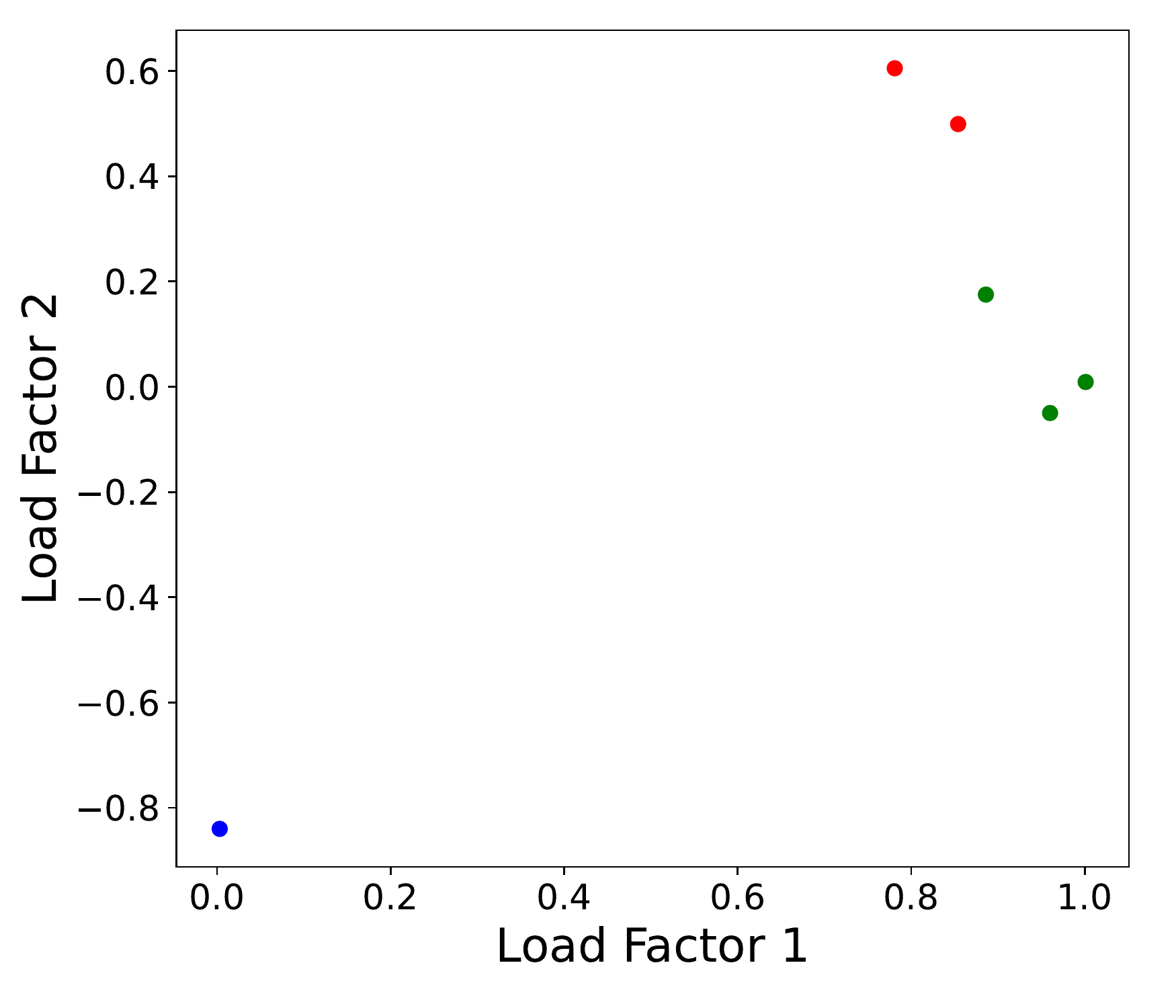}  
\caption{Factor loadings of the variables for activity (blue), rotation
  and cycle periods (red), and colour, convection zone depth and convective
  turnover time (green).}  
\label{load_factor}  
\end{figure}
disposal the measurements of the rotation period and cycle periods, the stellar
colour, depth of the convection zone, and convective turnover time, as well
as the activity as measured by $R^+_{HK}$  (i.e. six parameters for each
star). Factor analyses then attempt to describe six corresponding
variables $X_1,...X_6$ for each star in terms of a smaller number of
variables $Y_1,...Y_k$. In a first step, it is determined how many of
these latent variables are in fact required to improve the residuals;
for all of our analyses, we used the FactorAnalyzer package in Python.

We used normalised variables and verified  that the correlation matrix of the data is not
equal to the identify matrix using Bartlett's test for
sphericity; we then used the Kaiser-Meyer-Olkin (KMO) 
criterion to assess whether the given data is suitable for factor analysis. 
We then proceeded to determine the eigenvalues and eigenvectors of the
correlation matrix. Applying the KMO criterion, we find two eigenvalues
in excess of unity and  we may therefore assume that two latent variables exist.

With the corresponding eigenvectors, we can proceed to compute the factor
loadings of the six original variables in terms of the two latent variables;
the result is shown in Fig.~\ref{load_factor}.
We can recognise that the first loading factor provides large loadings for
the variables colour, convection zone depth, and convective turnover time, while
the second factor provides large loadings for the variables for activity as well as the
rotation and cycle periods. It is therefore recommend that one of the
two latent variables be associated with the property of 'activity', the other one
with the property of 'stellar structure'.

These results are corroborated by a principal component analysis,
which shows that the introduction of a second variable, for example,
the $B-V$ colour index, to describe the relation $P_{\rm{cyc}}$ versus
$P_{\rm{rot}}$ reduces the total data variance by $\sim10\%$ and is, hence, justified. 

\begin{figure}  
\centering  
\includegraphics[scale=0.45]{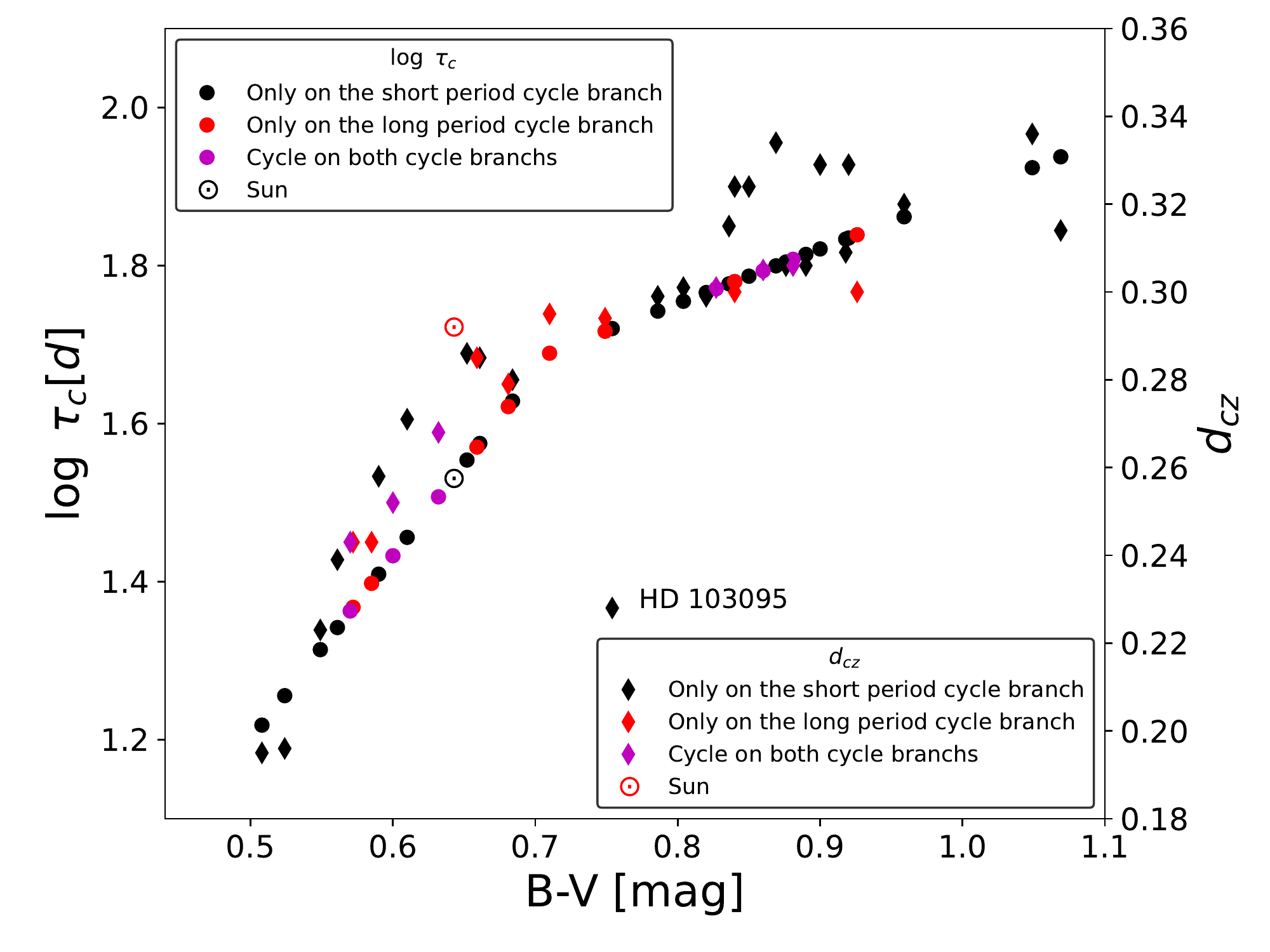}  
\caption{Convective turnover time ($\tau_{c}$) and the
relative depth of the convection zone ($d_{cz}$) versus $B-V$ [mag]; 
the data for $\tau_{c}$ are a labelled as dots and $d_{cz}$ as diamonds;
the colour-coding of the data indicates on which branch the corresponding
cycle is located.}
\label{tau_relative_depth-bv}
\centering
\end{figure}

We finally note that for main sequence stars, both the convective turnover time
($\tau_{c}$) and the relative depth of the convection zone ($d_{cz}$),
are strongly correlated with the stellar colour, as demonstrated in
Fig.~\ref{tau_relative_depth-bv}, where $\tau_{c}$ and $d_{cz}$ are
plotted vs. $B-V$ colour index for our sample stars. We note that the
$B-V$ colour is quite similar for both parameters; the only outlier
is the star HD~103095, which has a very low metalicity that leads to
an opacity and, hence, the depth of convection zone is different from that of the
other stars.

\subsubsection{Cycle-rotation period relation with unity slope and
  additional parameters}

In the following, we describe our attempts to improve the empirical
cycle-rotation period relations by considering further parameters and to check
the residuals from the derived cycle period expectations against  the
observed stellar cycle periods.
For this purpose, we start again (following the theoretical expectations outlined in
Sec.~\ref{theory}) by assuming a direct proportionality between the logarithmic
cycle and rotation periods with slope unity, but now improving it with an
additional parameter, namely: the $B-V$ colour index, the relative depth of
the convection zone, and the convective turnover time.

\begin{figure}  
\centering  
\includegraphics[scale=0.5]{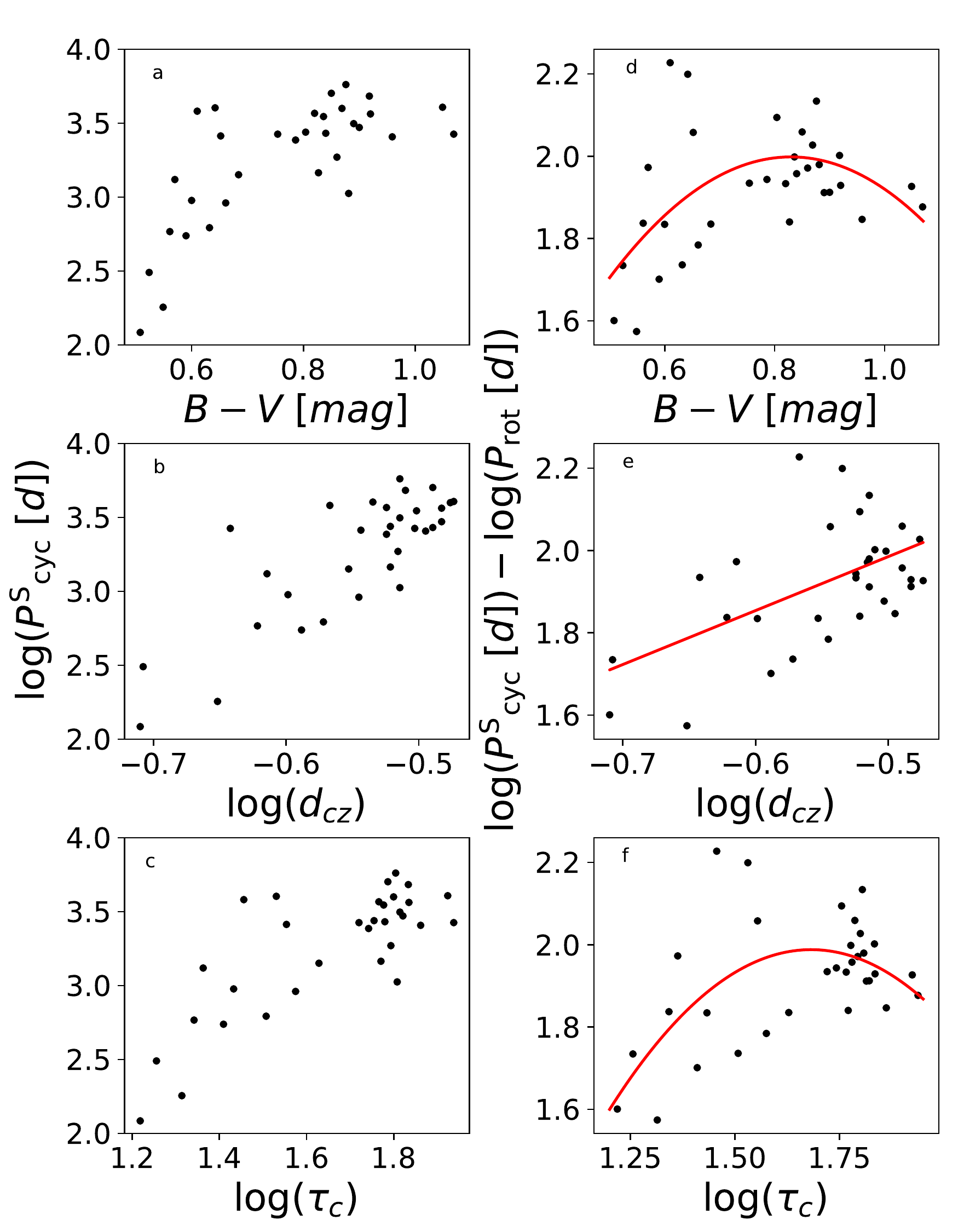}
\caption{Correlation between logarithm of the cycle period and
the colour index $B-V$, relative depth of the convection zone and
  convective turnover time. Left panels: Logarithm
  of the cycle period
  versus the colour index $B-V$ in panel a,
  versus the relative depth of the convection zone in panel b, and 
  versus the convective turnover time in
  panel c.
  Right panels:\ Difference between the logarithm
  of the cycle period and the logarithm of the
  rotation period versus the colour index $B-V$ is shown in panel d,
  versus the relative depth of the convection zone  in panel e and
  the convective turnover time in panel f. The red solid line in
  panels d to e shows the estimated trend between both values; see text for details.}
\label{regression_dr_tau_bv}
\centering
\end{figure}

We first considered the empirical dependence of the observed cycle
periods on these three parameters; for an illustration, see the plots shown in
Fig.~\ref{regression_dr_tau_bv}(a-c). As is evident to the eye, a clear
correlation between the cycle period and each of these three parameters
exists. To remove any indirect effects  of the rotation period, we considered
the quantity $log(P_{\rm{cyc}}/P_{\rm{rot}})$ and studied its dependence on those
three parameters; Fig.~\ref{regression_dr_tau_bv}(d-f) gives an idea of
these dependences and demonstrates a clear correlation between the
logarithmic period differences and the chosen parameters.
By removing the effect of the  rotation, the correlation between $log(P_{\rm{cyc}}/P_{\rm{rot}})$
and those three parameters becomes smaller when compared to the correlation between $log(P_{\rm{cyc}})$ and
those three parameters, which is also visible in Fig.~\ref{regression_dr_tau_bv}. Nevertheless,
the correlations between $log(P_{\rm{cyc}}/P_{\rm{rot}})$
and those three parameters have a significance of at least of 95$\%$.
In the following, we discuss these dependences individually.

\myparagraph{ $B-V$ colour index}
\label{cyc_prot_bv}

We first considered a direct dependence of  $log(P_{\rm{cyc}}/P_{\rm{rot}})$ on
the  $B-V$ colour index.  Figure~\ref{regression_dr_tau_bv}(d) suggests
a quadratic dependence and a least-squares analysis results in the
best-fit expression:
\begin{eqnarray}
  \label{eq_per_bv_res}
  \log(P^{S}_{cyc}) & = & (1.50\pm0.12) + \log(P_{rot}) \\ \nonumber
  & & + (2.32\pm0.75) X + (-2.7\pm1.0)X^{2},
\end{eqnarray}
where we define $X = B-V-0.4$, In Fig.~\ref{regression_dr_tau_bv}d, the curve 
described by Eq.~\ref{eq_per_bv_res} is plotted with the red line. 

\myparagraph{Relative depth of the convection zone}
\label{cyc_prot_relative_depth}

We now assume a direct proportionality between $log(P_{\rm{cyc}}/P_{\rm{rot}})$
and the (logarithmic) relative depth of the convection zone, and (again) a least-squares analysis results in the best-fit expression:
\begin{eqnarray}
  \label{eq_per_dr_res}
\log(P^{S}_{cyc}) = (2.64\pm0.20) + \log(P_{rot}) + (1.31\pm0.36) \log(d_{cv}),
\end{eqnarray}
with  $d_{cz}$ denoting the relative depth of the convection zone.
In Fig.~\ref{regression_dr_tau_bv}, the curve described by
Eq.~\ref{eq_per_dr_res}e is again plotted with the  red line. 

\myparagraph{Convective turnover time}
\label{cyc_prot_tau}

Finally, we compared $log(P_{\rm{cyc}}/P_{\rm{rot}})$ and
the (logarithmic) convective turnover time. From
Fig.~\ref{regression_dr_tau_bv}(f), we assume a quadratic dependence
and a least-squares analysis results in the best-fit expression:
\begin{eqnarray}
  \label{eq_per_tau_res}
  \log(P^{S}_{cyc}) & = & (1.60\pm0.09) + \log(P_{rot}) \\ \nonumber
  &  & + (1.61\pm0.51)Y + (-1.67\pm0.65)Y^{2},
\end{eqnarray}
where we define $Y=\log(\tau_{c})-1.2$ and $\tau_{c}$ denotes
the convective turnover time (in days).
In Fig.~\ref{regression_dr_tau_bv}f, the curve described by
Eq.~\ref{eq_per_tau_res} is again plotted with the  red line. 

With the introduction of the empirical convective turnover time in the
cycle-rotation relation, we can substitute the rotation period with
the above-introduced Rossby number, which is the ratio between the rotation
period and the convective turnover time.
The obtained relation (see Eq.~\ref{eq_per_tau_res}) can then be rewritten
in the form: $P^{S}_{cyc} \propto  Ro$.

\subsubsection{Double logarithmic cycle-rotation relation with variable slope and a $B-V$ splitting}
\label{cyc_per_bv_ranges}

In this subsection, we analyse how much it is possible to improve the empirical cycle-rotation relations with
a variable slope (i.e. not fixed at unity) by considering
the possible $B-V$ colour index dependence. 
\begin{figure}  
\centering  
\includegraphics[scale=0.45]{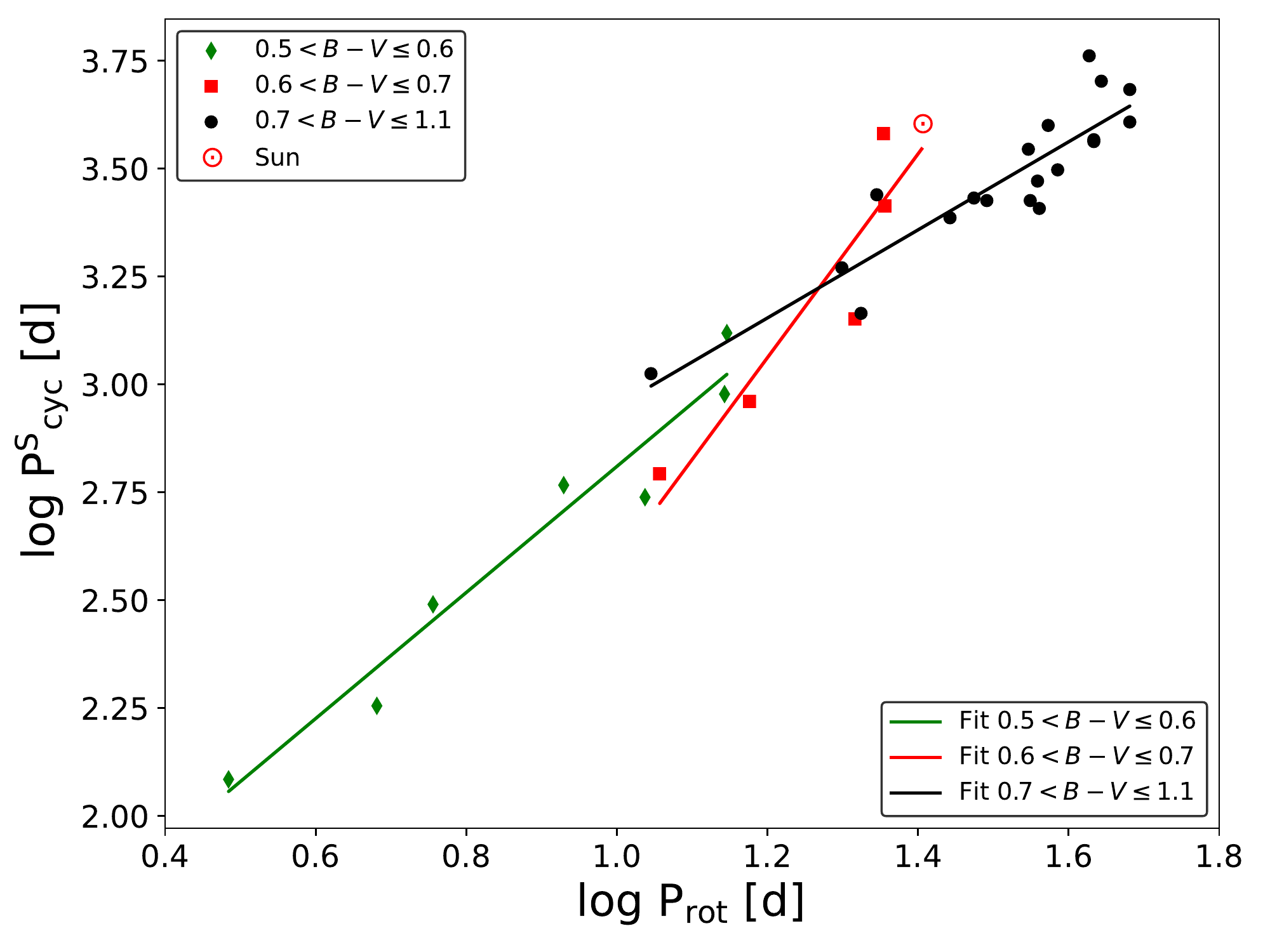}  
\caption{Logarithms of short-cycle branch periods are
  plotted over the logarithm of rotation period. Different $B-V$ ranges
  are colour coded and the solid lines depict the
  estimated empirical relations of individual $B-V$ ranges, which discussed in Sec. \ref{cyc_per_bv_ranges}.}  
\label{cyc_vs_prot}  
\end{figure}

To check the $B-V$ dependence, we split the data into different $B-V$ ranges,
labelled by different
colours and symbols in Figs. \ref{bv_plot}  and \ref{cyc_vs_prot}.
We also point out (in Sect. \ref{per_rot_rossby_rel} and
Fig. \ref{log_cyc_vs_log_ro}) a clear splitting into these $B-V$
ranges. From an inspection of Fig.~\ref{cyc_vs_prot}, it is obvious that different
spectral types cover different period ranges. 
Next, we estimate the trends between $\log P^{S}_{cyc}$ and $\log P_{rot}$
with the linear model (see Eq. \ref{lin_log_pcyc_log_ro_a}) through a least-squares fit in those three $B-V$ ranges.
The results are listed in Table~\ref{tab_trend_test},
including the number of data points in each $B-V$ range. This listing reveals
differences between the individual $B-V$ ranges. For the later spectral types, the derived best-fit slopes are indeed near unity,
whereas the slopes become greater than unity for the earlier spectral types,
suggesting again that the trend between the cycle period and the
rotation period depends on spectral type and, thus, on $B-V$.

Next, we computed the relative deviation
($\Delta$P$_{\rm{cyc}}^{S}/$P$_{\rm{cyc}}^{S}$) to check the
reduction of the remaining scatter compared to the cycle-period
relation (see Eq. \ref{pcyc-prot}) and obtained a standard
deviation ($\sigma$) of 0.20, namely,  a reduction in the scatter
of $\approx$15$\%$.

Finally, we calculated the solar cycle period with the obtained
relation for the $B-V$ range from 0.6 to 0.7, listed in Table \ref{tab_trend_test}, and we obtained a cycle
period of 9.6$^{+31.5}_{-7.4}$ yr. However, due to the large error in the slope, the 
error margins of this value are extremely large, thus making this cycle period very uncertain.
Nevertheless,  this calculated period is consistent with the well-known 11-year Schwabe cycle and it is
located in the range of the measured cycle lengths for the Schwabe cycle.

\begin{table}[!t]  
  \caption{Results of the $B-V$ trend estimation to test a possible secondary $B-V$
    dependence (see Fig.~\ref{cyc_vs_prot})}  
\label{tab_trend_test}  
\begin{center}  
\begin{small}  
\setlength{\tabcolsep}{5pt}
\begin{tabular}{lcccc}
\hline  
\hline  
\noalign{\smallskip}  
$B-V$ range & No. of & a & n & $\sigma$ of  \\
 & data & & & $\Delta$P$_{\rm{cyc}}^{S}/$P$_{\rm{cyc}}^{S}$ \\
\hline  
\noalign{\smallskip}  
$0.5<B-V\leq0.6$  & 7 & 1.35$\pm$0.05 & 1.46$\pm$0.08 & 0.20\\
$0.6<B-V\leq0.7$  & 6 & 0.23$\pm$0.13 & 2.36$\pm$0.44 & 0.29\\
$0.7<B-V\leq1.1$  & 19 & 1.93$\pm$0.06 & 1.02$\pm$0.04 & 0.17\\
\hline  
\end{tabular}
\end{small}  
\end{center}  
\end{table}

\begin{figure}  
\centering  
\includegraphics[scale=0.5]{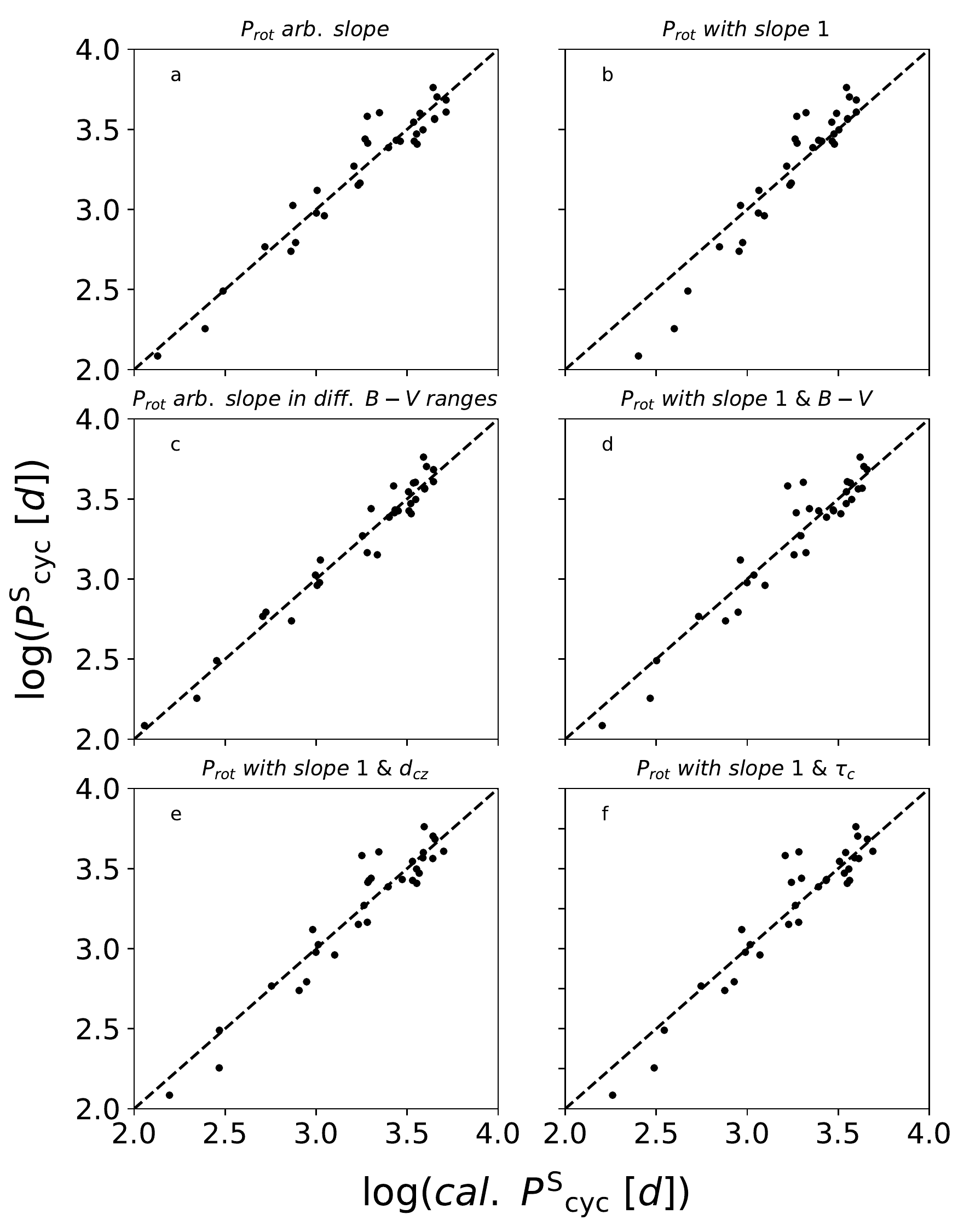}  
\caption{ Measured versus  calculated
  cycle period as the basis of the comparison among the different models. Panel a:
  Linear model (Eq.~\ref{lin_log_pcyc_log_ro_a}) with a best-fit
  slope for $\log(P_{rot})$.
  Panel b: Linear model (Eq.~\ref{lin_log_pcyc_log_ro_a})
  with a slope of 1 for $\log(P_{rot})$. Panel c: Linear model
  (Eq.~\ref{lin_log_pcyc_log_ro_a}) with a best-fit slope for $\log(P_{rot})$
  for different $B-V$ ranges. Panel d: Model (Eq.~\ref{eq_per_bv_res})
  with a slope of 1 for $\log(P_{rot})$ and $B-V$ term. Panel e: Model (Eq.~\ref{eq_per_dr_res}) with a slope of 1
  for $\log(P_{rot})$ and the relative depth of the convection zone. Panel f: Model (Eq.~\ref{eq_per_bv_res}) with a slope of 1 for
  $\log(P_{rot})$ and the convective turnover time. }  
\label{reg_analysis}  
\end{figure}

\subsection{Summary of the different approaches to the
  $\log~P^{S}_{cyc}$ versus $\log~P_{rot}$ relation}
\label{sum_diff_approaches}
In our regression analyses, we considered different models that make different
assumptions on the proportionality between the logarithmic cycle and
rotation periods and additional stellar parameters.   In this subsection, we
compare how well the different empirical representations reproduce the
actual observed cycle periods (cf., Fig.~\ref{reg_analysis}).
To quantify the `goodness of fit' of any empirical relation by its residuals,
we used the relative differences between observed and
measured periods, listed in Table~\ref{tab_reg_analysis}.

\begin{table}[!t]  
  \caption{The relative deviations of the different $P^{S}_{cyc}$ versus $P_{rot}$
    models}  
\label{tab_reg_analysis}  
\begin{center}  
  \begin{small}
    \scalebox{1.}{
\begin{tabular}{lcc}  
\hline  
\hline  
\noalign{\smallskip}  
Model & Fig.~\ref{reg_analysis} &  $\sigma$ of $\Delta$P$_{\rm{cyc}}^{S}/$P$_{\rm{cyc}}^{S}$ \\
& panel & \\
\hline  
\noalign{\smallskip}
$P_{rot}~arb.~slope$ & a & 0.24\\
$P_{rot}~with~slope~1$ & b & 0.40 \\
$P_{rot}~arb.~slope~in~diff.~B-V~ranges$ & c & 0.20\\
$P_{rot}~with~slope~1$ \& $B-V$ & d & 0.27 \\
$P_{rot}~with~slope~1$ \& $d_{cz}$ & e & 0.28 \\
$P_{rot}~with~slope~1$ \& $\tau_{c}$ & f & 0.29 \\
\hline  
\end{tabular}
}
\end{small}  
\end{center}  
\end{table}

An inspection of Table~\ref{tab_reg_analysis} shows that the best-fit slopes
(hence, different from unity) in the cycle-rotation relation already yield
 a very good description, which can be further improved upon
by splitting up the data into different colour regimes.
However, enforcing a slope of unity always leads to 
poorer results, even though they can still be improved by distinguishing
between different $B-V$-colours, the relative depth of the convection
zone, or convective turnover time (all functions of $B-V$-colour).
However, none of these additional considerations allows for a description
that produces the same fit quality of those obtained with the best-fit slopes.
\begin{table}[!t]  
\caption{Results of the overall F-tests}  
\label{tab_ftest_analysis}  
\begin{center}  
\begin{small}
\begin{tabular}{lcc}  
\hline  
\hline  
\noalign{\smallskip}  
Model & F-value &  p-value \\
\hline  
\noalign{\smallskip}
\multicolumn{3}{c}{compare to $P_{rot}~with~slope~1$} \\
\hline  
\noalign{\smallskip}
$P_{rot}~with~slope~1$ \& $B-V$ & 2.81 & 0.0032 \\
$P_{rot}~with~slope~1$ \& $d_{cz}$ & 2.53 & 0.0063 \\
$P_{rot}~with~slope~1$ \& $\tau_{c}$ & 3.81 & 0.0003 \\
\hline
\multicolumn{3}{c}{compare to $P_{rot}~with~slope~1$} \\
\hline  
\noalign{\smallskip}
$P_{rot}~arb.~slope$ & 5.18 & 9.90$\cdot10^{-6}$ \\
\hline
\multicolumn{3}{c}{compare to $P_{rot}~arb.~slope$} \\
\hline  
\noalign{\smallskip}
$P_{rot}~arb.~slope~in~diff.~B-V~ranges$ & 4.31 & 0.0002\\
\hline  
\end{tabular}
\end{small}  
\end{center}  
\end{table}

To test for the significance of the reduction of the residuals by use of
best-fit slopes compared to a reference model, we used the overall F-test
for either type of empirical relation. In this overall F-test, the different
numbers of degrees of freedom are considered, so that a comparison of the models
independent of the degrees of freedom is possible.
The resulting p-values give the confidence level of the reduction of the relative
deviations. In all cases, the obtained p-values indeed suggest  a significant
improvement over the relations based on unity slopes.

In summary, these overall F-tests suggest that the best empirical
representation of the cycle-rotation relation is a power law
(i.e. with a non-unity best-fit slope in the double-logarithmic diagram) 
when distinguishing between different $B-V$ ranges. The emerging
$B-V$ dependence has two possible reasons: while the $B-V$ dependence
of the cycle period is well known from the activity-rotation relation,
the second $B-V$ dependence (i.e. how much the cycle period actually depends on
rotation period) seems to be related to stellar structure.

\section{Rossby number as a natural parameter for the cycle-rotation relation}
\label{norm_cyc_ro}

In Sect. \ref{cyc_prot_tau}, we introduce the convective turnover
time in the cycle-rotation relation, which opens up the possibility
of using the Rossby number as the physical parameter,  instead of
the rotation period. In the landmark paper by
\citet{Noyes1984ApJ...279..763N}, a simple (colour-independent)
rotation-activity relation was found once the rotation periods
are `parameterised' by the Rossby number. Therefore, we 
decided to study the possible improvements coming from the use of the Rossby number.

Using the Rossby number to `scale' the actual rotation periods leads to a
shift onto a comparable rotation scale, irrespective of colour. 
To compute the Rossby numbers for our sample stars, we used the prescription for the
calculation of the empirical convective turnover time derived
in \citep{Mittag2018}; with this empirical definition of the convective
turnover time, a clear definition range between 0 and 1 for the
Rossby number \citep{Mittag2018} can be obtained. This approach is consistent
with dynamo modelling, which suggests that cyclic activity is obtained only
for Rossby numbers smaller than one, equivalent to a dynamo number larger
than unity.

Furthermore, taking into account the best empirical representation
above is of the form so that $P^{S}_{\rm{cyc}} \propto P_{\rm{rot}}^{n}$, we 
use the same type of ansatz:\ 
\begin{eqnarray} \label{stix_cycle_freq_rewrite_d}
  P^{S}_{\rm{cyc}} & \propto & A \times Ro^{n},
\end{eqnarray}
where the constant, $A,$ denotes some timescale. We used the
convective turnover time, $\tau_c$, as the timescale for
this constant, $A$.

\subsection{Empirical $P^{S}_{\rm{cyc}}-Ro$ relation}
\label{per_rot_rossby_rel}

\begin{figure}  
\centering  
\includegraphics[scale=0.45]{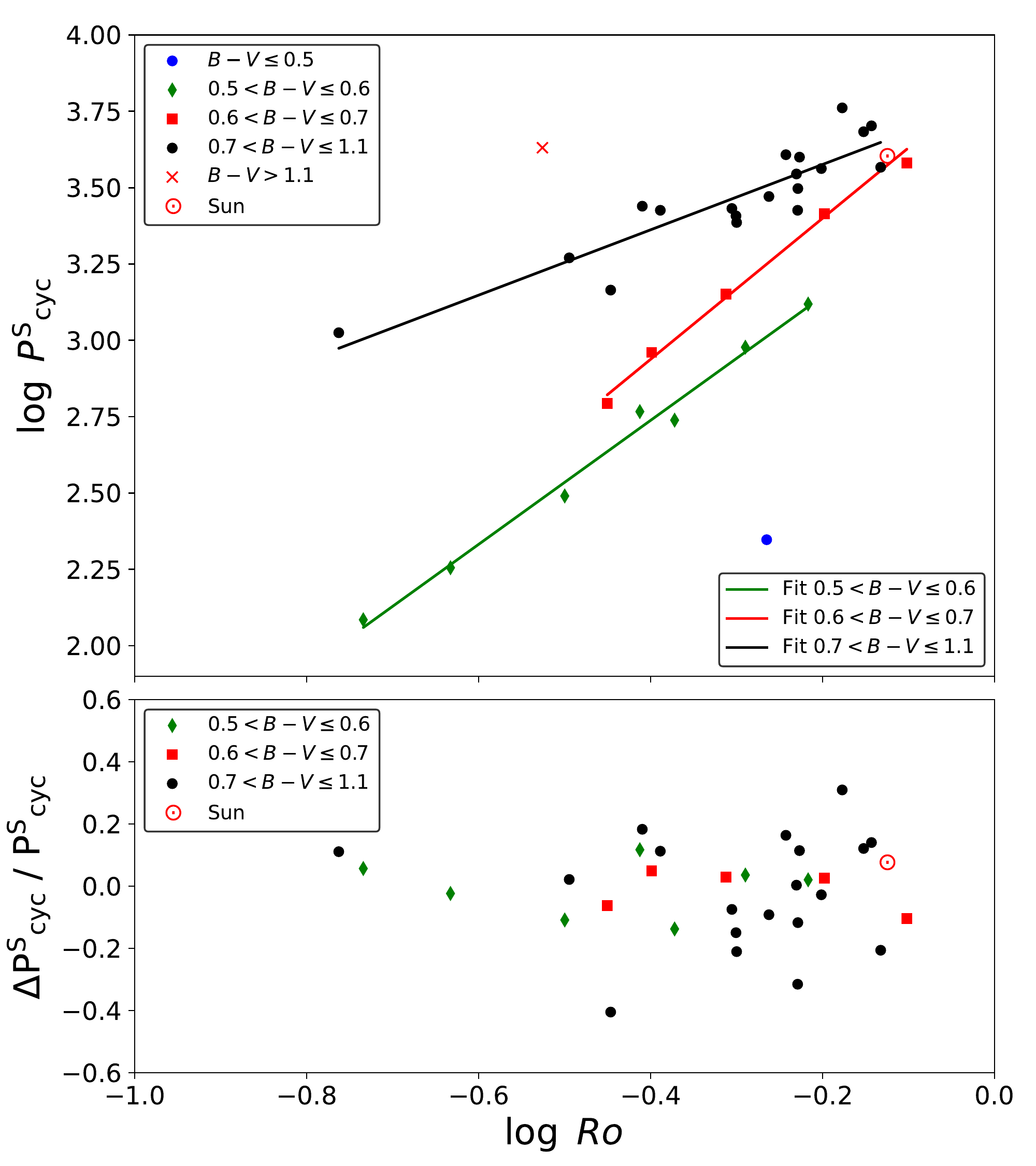}  
\caption{Logarithm of cycle periods on the short-cycle branch
  versus the logarithm of the Rossby number. Upper panel: 
  Cycle period against Rossby number for cycles on the short-period
  (inactive) branch; different colours and symbols indicate 
  the different $B-V$ ranges. The solid-coloured lines show the respective
  empirical relations, with colours indicating the different $B-V$
  ranges. Lower panel: Relative deviation
  ($\Delta P^{\rm{S}}_{\rm{cyc}}/P^{\rm{S}}_{\rm{cyc}}$) 
  versus the logarithm of the Rossby number.  The green points label
  the stars within the $B-V$ range of $0.5<B-V\leq 0.6$, the red points
  stars within $0.6<B-V \leq 0.7$, and the black points those within
  $0.7<B-V\leq 1.1$.}  
\label{log_cyc_vs_log_ro}  
\end{figure}

In the upper panel of Fig. ~\ref{log_cyc_vs_log_ro},  we plot  the cycle periods versus the
corresponding Rossby numbers for our sample stars in a double logarithmic scale; it
is clear that the introduction of the Rossby number leads to a
horizontal shift of the data points shown in Fig.~\ref{cyc_vs_prot},
which then changes the simple cycle-rotation period relation described by
Eq.~\ref{pcyc-prot}. However, a re-scaling of the rotation periods
with the empirical convective turnover time leads to an almost identical
Rossby number range for the different stellar samples, which serves as an argument in favour
of this approach. Furthermore, the upper panel of Fig. ~\ref{log_cyc_vs_log_ro} suggests
that the cycle periods are divided into three main $B-V$ ranges, $0.5 < B-V \le 0.6$, $0.6 < B-V \le 0.7$
and $0.7 < B-V \le 1.1.$  Therefore, we used this $B-V$ splitting for our study in
Sect. \ref{cyc_per_bv_ranges} and labelled these ranges in Figs.~\ref{bv_plot} and \ref{cyc_vs_prot}.
Also, the trends between the cycle periods and Rossby numbers in these
$B-V$ ranges  differ; two data points are clearly
outside of these $B-V$ ranges and these two points are not considered further in our study; thus, the
numbers of the used cycle periods is reduced from 34 to 32 cycles.

We then fit the data in the three $B-V$ ranges using the logarithmic
form of Eq.~\ref{stix_cycle_freq_rewrite_d} and a simple linear least-squares fit that takes the form: 
\begin{eqnarray}\label{lin_log_pcyc_log_ro}
  \log P^{S}_{\rm{cyc}} & = & \log(A_{i})+n^{S}_{i} \log Ro, 
\end{eqnarray}
where the index $i$ labels the different $B-V$ ranges. Thus, we obtain the results listed in Table \ref{tab2}.
Also, the derived regression curves are shown as solid coloured lines
in Fig.~\ref{log_cyc_vs_log_ro}.  

A comparison of the slopes in Table~\ref{tab_trend_test} shows that these
are similar within the error margins, except for the colour range
of $0.5<B-V\leq0.6$. Similarly to the results obtained in Sect. \ref{cyc_per_bv_ranges},
these findings again suggest a $B-V$ dependence of the trends. 
We note that for the stars of earlier spectral
types, the derived slopes are not consistent with unity, while for the later
types, they are consistent once the Rossby number is used as a parameter.
\begin{table}[!t]  
  \caption{Results of the $B-V$-specific empirical relations (see Fig.~\ref{log_cyc_vs_log_ro})}  
\label{tab2}  
\begin{center}  
  \begin{small}
    \setlength{\tabcolsep}{5pt}

\begin{tabular}{lcccc}  
\hline  
\hline  
\noalign{\smallskip}  
B-V range  & No. of & $\log(A)$ & $n^{S}$ & $\sigma$ of \\
            &  data &         &          & $\Delta$P$_{\rm{cyc}}^{S}/$P$_{\rm{cyc}}^{S}$ \\
\hline  
\noalign{\smallskip}  
$0.5<B-V\leq0.6$  & 7 & 3.55$\pm$0.05 & 2.03$\pm$0.09 & 0.09 \\
$0.6<B-V\leq0.7$  & 6 & 3.86$\pm$0.03 & 2.31$\pm$0.10 & 0.07 \\
$0.7<B-V\leq1.1$  & 19 & 3.79$\pm$0.04 & 1.07$\pm$0.13 & 0.19\\
\hline  
%\tableline  
\end{tabular}
\end{small}  
\end{center}  
\end{table}  

Using the fit results displayed in Table~\ref{tab2}, we can again calculate
the relative deviations ($\Delta P^{\rm{S}}_{\rm{cyc}}/P^{\rm{S}}_{\rm{cyc}}$),
plotted in the lower panel of Fig.~\ref{log_cyc_vs_log_ro}. The three $B-V$ ranges are
again colour-coded, green points denote stars
in the $B-V$ range of $0.5<B-V\leq0.6$, red points  stars in $0.6<B-V\leq0.7$,
and black points those in $0.7<B-V\leq1.1$.

To evaluate the scatter of
$\Delta P^{\rm{S}}_{\rm{cyc}}/P^{\rm{S}}_{\rm{cyc}}$, we calculated
the standard deviation of $\Delta P^{\rm{S}}_{\rm{cyc}}/P^{\rm{S}}_{\rm{cyc}}$ and
obtained a scatter value of 0.15 for all stars, implying an average deviation
of 15 $\%$ for the used stars. Compared to the remaining scatter for the
relation Eq.~\ref{pcyc-prot}, we see the residual deviation  is clearly
decreased by the use of the Rossby number as parameter.

Further support for this approach comes from the solar 11-year cycle.
Using the stellar relation that is valid for the solar $B-V$ colour range, we
find an expected cycle period of 10.2$^{+0.8}_{-0.7}$~yr, which
indeed nicely matches the 11-year Schwabe cycle.
Hence, in the framework of the  empirical cycle-rotation relation presented here,
the solar 11-year Schwabe cycle follows the trend between the
$\log P^{S}_{cyc}$ and $\log Ro$ very well. Consequently, in this relation, the
solar 11-year Schwabe cycle is no longer an outlier as 
in the relation between the $\log P^{S}_{cyc}$ and $\log P_{rot}$ (cf.  Fig. \ref{bv_plot}).

\subsection{Physical interpretation  of the $P^{S}_{\rm{cyc}}-Ro$ relation}
\label{constant_a_est}
Up to this point, we have look at the relationship between the cycle period
and Rossby number on purely empirical grounds. In our factor analysis (described in
Sect.~\ref{FA_PCA}), we found that the convective turnover time and the
relative depth of the convection zone can be viewed as latent variables in the
cycle-rotation relation. Furthermore, in
Sects.~\ref{cyc_prot_tau}~and~\ref{cyc_prot_relative_depth}, we consider
the significance of these additional parameter in the cycle-rotation
relation, with the exponent of the rotation period fixed at unity.
Therefore, we may assume that these hitherto unconsidered parameters are
to be found again in the constant A in Eq.~\ref{stix_cycle_freq_rewrite_d}. 

\begin{figure}  
\centering  
\includegraphics[scale=0.5]{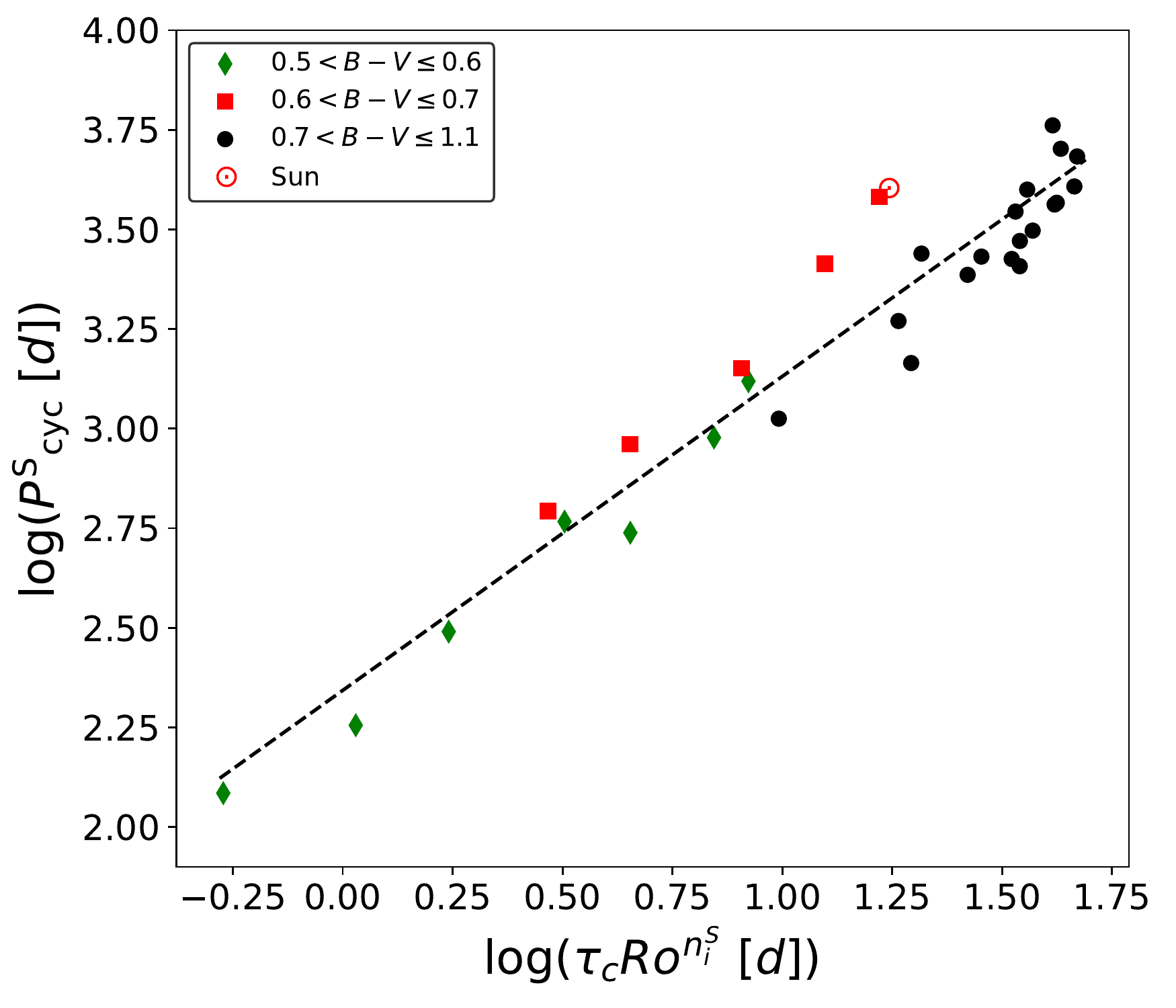}
\caption{Comparison of $\log P^{S}_{\rm{cyc}}$ with $\log(\tau_{c}Ro^{n^{S}_{i}})$,
    where the $B-V$ ranges are colour-coded.}  
\label{cyc_per_short_vs_tau_c-rosby}
\end{figure}

To test the effects of the introduction of
$\tau_{c}$ on the $P^{S}_{\rm{cyc}}-Ro^{n^{S}_{i}}$ relation, 
we compared ( in Fig. \ref{cyc_per_short_vs_tau_c-rosby})
the values of $\log P^{S}_{\rm{cyc}}$
with $\log(\tau_{c}Ro^{n^{S}_{i}})$, with the coefficients $n^{S}_{i}$ listed in Table~\ref{tab2}.  
It is evident that
that the splitting between different B-V ranges (as seen in Fig. \ref {log_cyc_vs_log_ro}) has mostly
disappeared and the relation between
between $\log P^{S}_{\rm{cyc}}$ and $\log(\tau_{c}Ro^{n^{S}_{i}})$ is
linear. However, it is also visible that the different $B-V$ ranges are slightly shifted 
around the mean trend, with the cooler stars lying mostly below and the hotter stars lying
mostly above the regression line.
We assume that these effects are caused by the hitherto unconsidered length scale of the
turbulence (cf., Eq.~\ref{stix_cycle_freq_rewrite}), and,
furthermore, we assume that this turbulence length scale 
($l/R_{\star}$) is representative for the relative depth of the convection zone
where the dynamo is located. Assuming
that the dynamo is located at the bottom of the convection zone, we may use
$d_{cz}$   instead of $l/R_{\star}$. 
Finally, we remark that HD~103095 shows a strong deviation in the relative depth of the
convection zone-colour relation when compared to the other stars (see
Fig.~\ref{tau_relative_depth-bv}); therefore, we excluded this star from this analysis.
  
\begin{figure}  
\centering  
\includegraphics[scale=0.5]{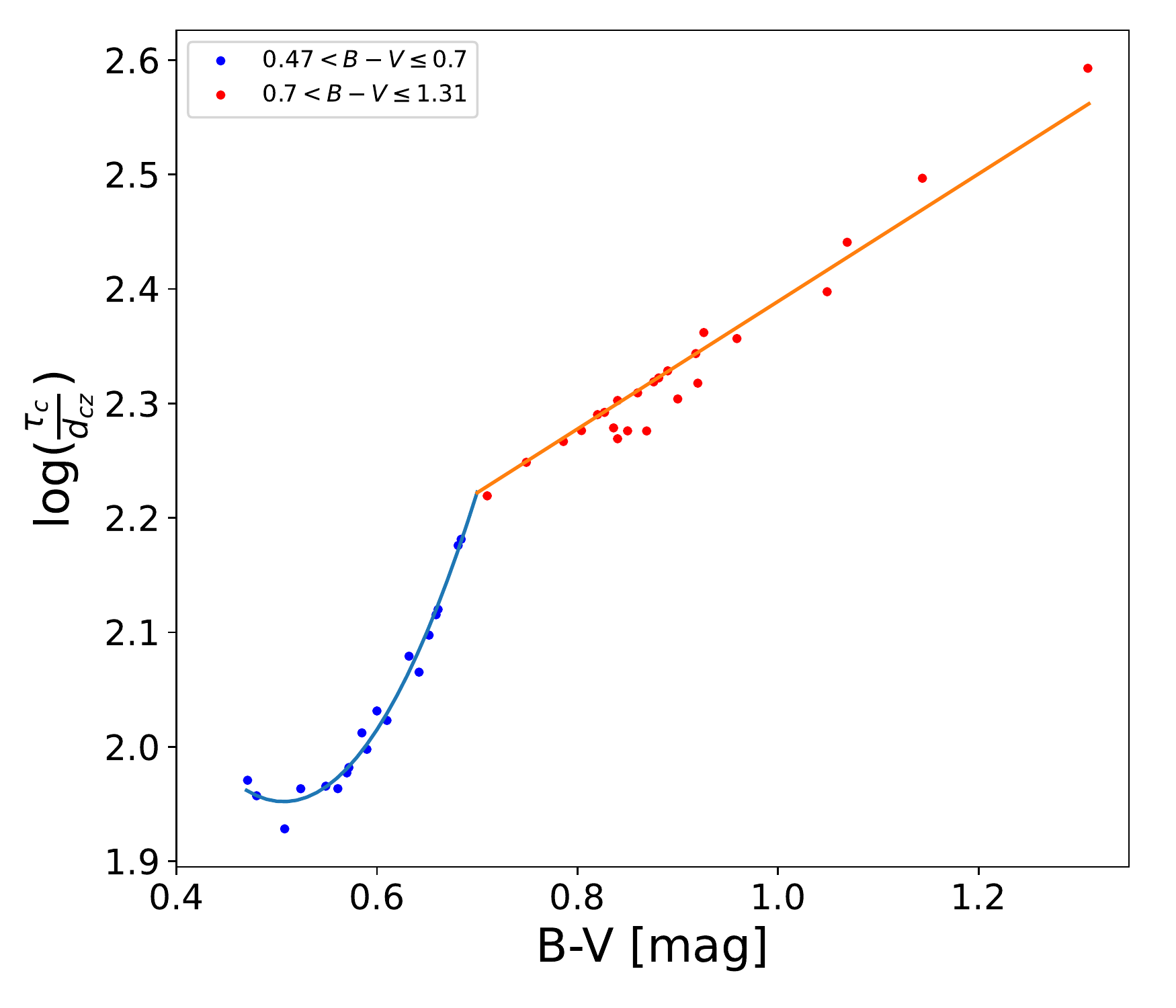}  
\caption{Logarithm of the ratio of the convective turnover time and relative depth of the
  convection are shown. Solid-coloured line depicts the estimated trend.
  Solid-blue line depicts the
  $B-V$ range of $0.47~<~B~-V~\le~0.7$ and the solid-red line shows the range
  of $0.7~<~B~-V~\le~1.31$.}  
\label{ratio_tau_reldcz}  
\end{figure}  

Accordingly, we assume that the ratio of $\tau_{c}$ and $d_{cz}$ is
contained in the constant $A$ introduced above and show this ratio of our sample stars
in Fig. \ref{ratio_tau_reldcz}.
Physically, this ratio acts as the inverse of velocity,
describing the speed of the convective overturn in a convection zone; naturally, this velocity
depends on $B-V,$ such that:
\begin{eqnarray} \label{definition_A}
A &  \sim & \left(\frac{\tau_{c}}{d_{cz}}\right) = v^{-1}_{dcz}(B-V). 
\end{eqnarray}
In Fig. \ref{ratio_tau_reldcz}, a clear change is apparent at the level of colour
$B-V\approx0.7$. Hence, we split the data at B-V = 0.7 and derived
two relations via a least-squares fit:
for the $B-V$ range $0.47 < B-V \le 0.7$, we obtain:

\begin{eqnarray} \label{rel_v_bv1}
  v^{-1}_{dcz} & = & (2.035\pm0.021) \nonumber \\
    & & -(1.55\pm0.24)X+(7.25\pm0.65)X^{2},
\end{eqnarray}

where $X = (B-V - 0.4)$, and for 
the $B-V$ range $0.7~<~B~-~V~\le~1.31$, we obtain: 

\begin{eqnarray} \label{rel_v_bv2}
  v^{-1}_{dcz} & = & (2.222\pm0.095) + (0.557\pm0.016)X,
\end{eqnarray}

where $X = (B-V-0.7)$. For this trend estimation, we also used
the four data points outside the  $B-V$ range $0.5 < B-V \le 1.1$.
The relations obtained in this way are depicted as a solid line in Fig. \ref{ratio_tau_reldcz}.
In the next section, we use these relation to derive our final $P^{S}_{\rm{cyc}}-Ro$ relation.

\subsection{Final $P^{S}_{\rm{cyc}}-Ro$ relation for the short-cycle branch}
\begin{figure}  
\centering  
\includegraphics[scale=0.5]{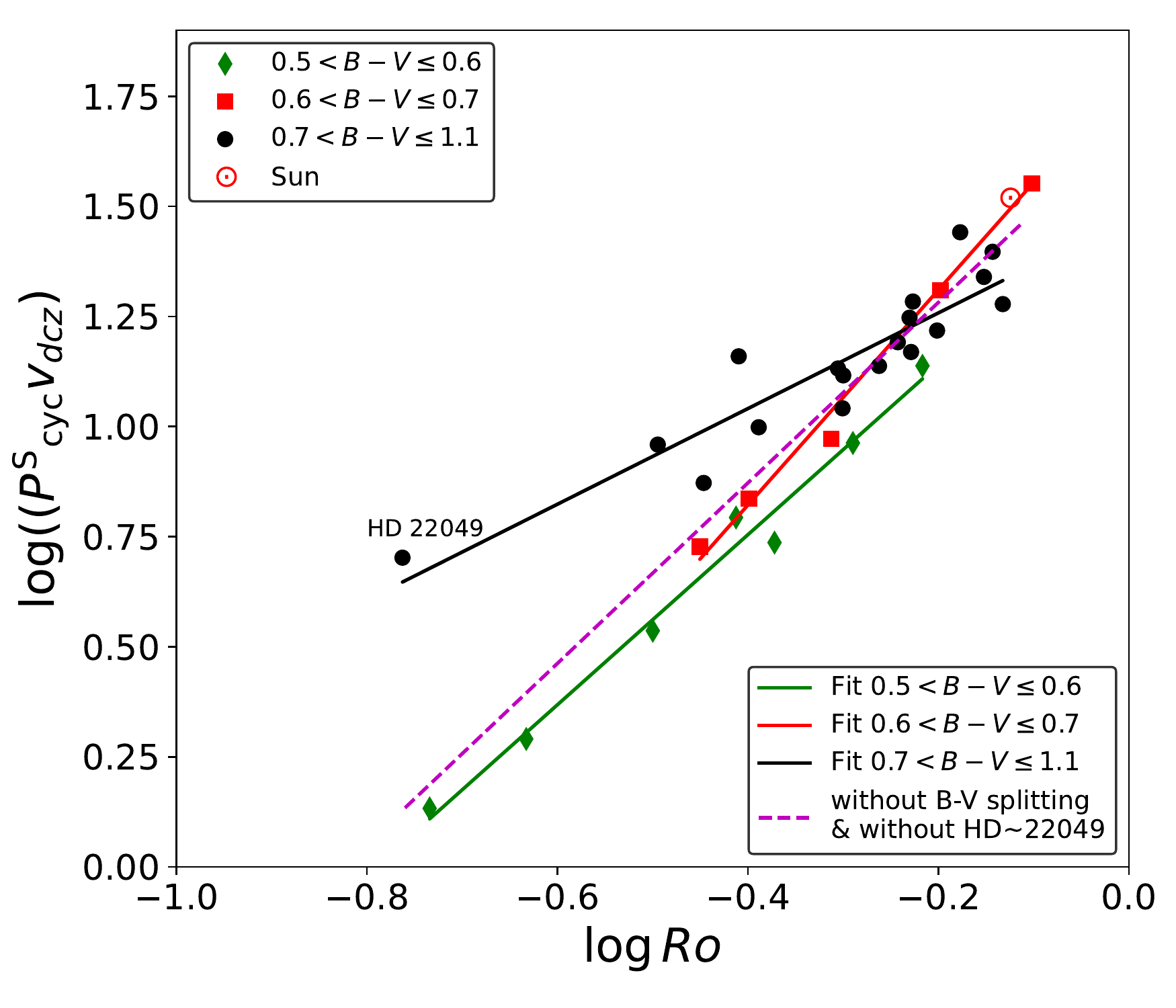}  
\caption{Ratio of the cycle period and constant $A$ (see Eq. \ref{definition_A})
  versus Rossby number in a double logarithmic scale, where the $B-V$ ranges
  are colour-coded. Furthermore, estimated trends
  in these $B-V$ ranges are depicted as solid and coloured lines.}  
\label{cyc_per_short_tau_c_dr_vs_rosby}
\end{figure}  

Using the relations in Eqs. \ref{rel_v_bv1} and \ref{rel_v_bv2},
we can compute the ratio of $\tau_{c}$ and $d_{cz}$ for all stars,
scale the corresponding cycle period with this value, and plot 
$\log(P^{S}_{\rm{cyc}} v_{dcz}(B-V))$ versus $\log(Ro)$
in Fig. \ref{cyc_per_short_tau_c_dr_vs_rosby} (again colour-coding the three main $B-V$ ranges).
Fig. \ref{cyc_per_short_tau_c_dr_vs_rosby} shows a linear relation between 
$\log(P^{S}_{\rm{cyc}} v_{dcz}(B-V))$ and $\log(Ro)$ with some `outliers'; 
alternatively, by introducing B-V ranges, we find  linear relations but with different
parameters. In the following, we study these relations using these approaches.

\subsubsection{Final fit without $B-V$ splitting}
\label{final_rel_with_bv_slit}
In Fig. \ref{cyc_per_short_tau_c_dr_vs_rosby}, we note a single blatantly outlying data point (at
$\log Ro \approx -0.76$ and $\log(P^{S}_{\rm{cyc}} v_{dcz}(B-V)) \approx 0.7$), 
which refers to HD~22049 (=eps Eri). 
This single data point of HD~22049 has a huge impact on our fit of the trend.
Considering this data point as outlier here and ignoring it in our least-squares  fit, we obtain:
\begin{eqnarray} \label{final_cycle_relation_short_without_bv}
  \log \left(\frac{P^{S}_{\rm{cyc}}d_{cz}}{\tau_{c}}\right) & = & \log \left(P^{S}_{\rm{cyc}} v_{dcz} \right) = \nonumber \\
  & & (1.692\pm0.049) + (2.05\pm0.14) \log Ro.
\end{eqnarray}
This result is plotted as magenta dotted line in Fig. \ref{cyc_per_short_tau_c_dr_vs_rosby}.
\begin{figure}  
\centering  
\includegraphics[scale=0.5]{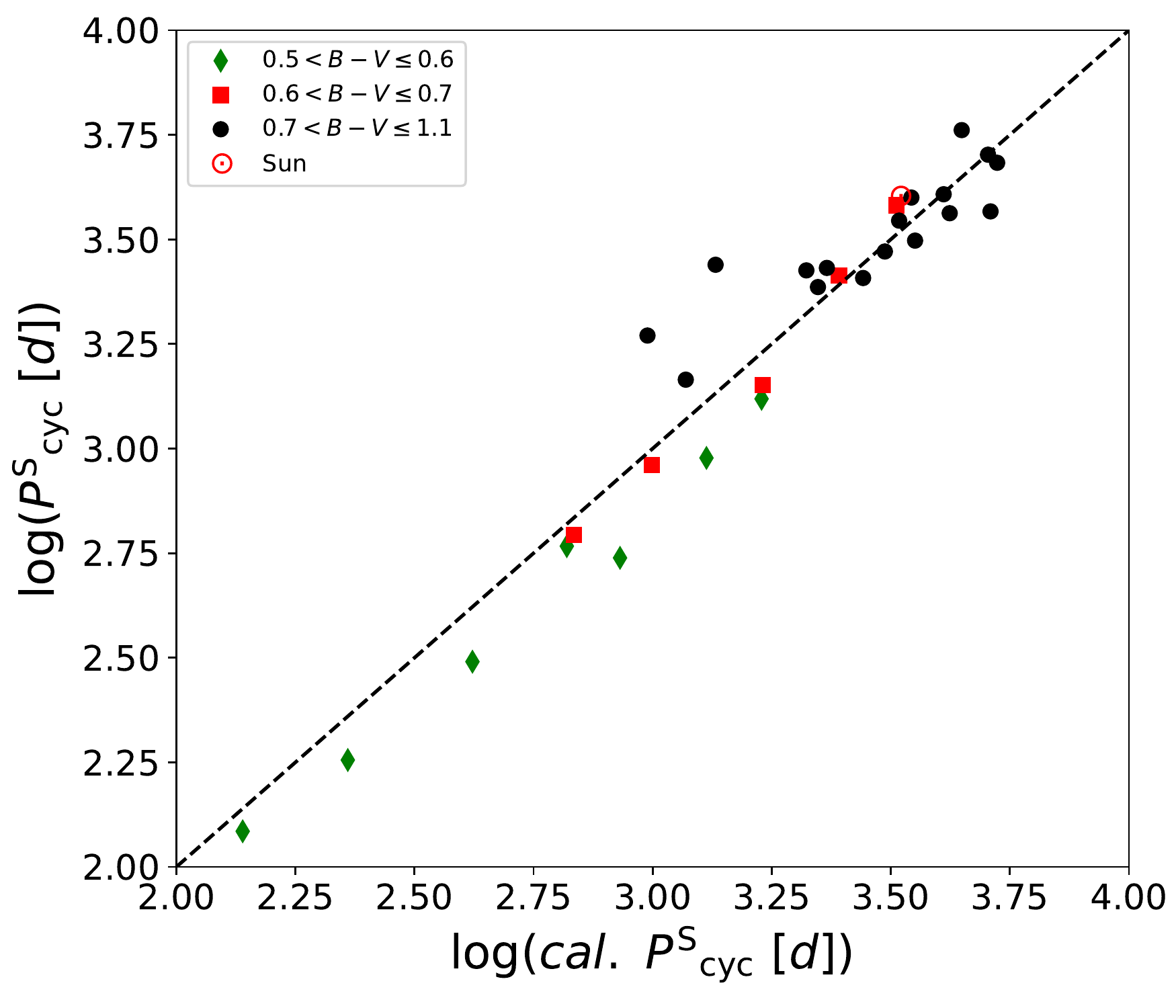}  
\caption{Comparison of the observed and calculated cycle periods from the relation of
  Eq. \ref{final_cycle_relation_short_A}, the $B-V$ ranges are colour-coded. The dashed line depicts the identity.}  
\label{cyc_per_short_vs_rosby_no_split}  
\end{figure}  

Next, we compare the observed and calculated cycle periods with the expression
\begin{eqnarray} \label{final_cycle_relation_short_without_bv_cal}
  P^{S}_{\rm{cyc}} & = & 10^{1.692\pm0.049} v^{-1}_{dcz} Ro^{2.05\pm0.14}
,\end{eqnarray}
and we plot the results in Fig. \ref{cyc_per_short_vs_rosby_no_split}.  Although
the three main B-V ranges in this trend estimation have not been considered, we colour-coded these
three $B-V$ ranges to check whether any possible $B-V$
dependencies or systematics remain. A inspection by eye of
Fig. \ref{cyc_per_short_vs_rosby_no_split} shows two possible cases where such $B-V$
systematics do indeed remain. First, it is clear that all data points for the $B-V$ range
$0.5<B-V\leq0.6$ are located below the regression curve which indicates that the calculated
cycle periods are systematically overestimated. Second, in the $B-V$ range
$0.7<B-V\leq1.1,$ the data seem to display a different trend.

To quantify the deviation between the observed and calculated cycle periods, we
determined the standard deviation of the relative deviations $\Delta P^{\rm{S}}_{\rm{cyc}}/P^{\rm{S}}_{\rm{cyc}}$ and
obtain 0.24. This value is similar the standard deviation of $\Delta P^{\rm{S}}_{\rm{cyc}}/P^{\rm{S}}_{\rm{cyc}}$
for the pure cycle-rotation relation (see Eq. \ref{pcyc-prot}). Thus, the relation 
in Eq. \ref{final_cycle_relation_short_without_bv_cal} does not provide a (statistically) better cycle-rotation connection
than pure cycle-rotation relation (see Eq. \ref{pcyc-prot}). However, for the Sun, we find
a cycle period of 9.0$^{1.2}_{1.0}$ yr close to the 11-year cycle
in contrast to the 6.1$^{+2.2}_{-1.6}$ yr cycle period estimated from Eq.~\ref{pcyc-prot}. 
Therefore, in this context the Sun is no longer an outlier, yet we have no explanation for the `strange' properties
of $\epsilon$ Eri.

\subsubsection{Final fit with $B-V$ splitting}
\label{final_cyc_ro_rel_sb}

Retaining all data points in the sample, it is not possible to find a single (linear) relationship;
rather, we find that a possible $B-V$ dependence remains.  As made clear from
Fig. \ref{cyc_per_short_tau_c_dr_vs_rosby}, in  the three main colour-coded $B-V$ ranges, 
the $\log(P^{S}_{\rm{cyc}} v_{dcz}(B-V))$ values show a linear dependence.
Assuming then a linear relation in each B-V range of the form:
\begin{eqnarray} \label{final_cycle_relation_short_A}
  \log \left(\frac{P^{S}_{\rm{cyc}}d_{cz}}{\tau_{c}}\right) = \log \left(P^{S}_{\rm{cyc}} v_{dcz} \right) & = & a^{S}_{i} + n^{S}_{i} \log Ro,
\end{eqnarray}
where the index $i$ labels the different $B-V$ ranges, we derived the parameters
$a^{S}_{i}$ and $n^{S}_{i}$ via a least-squares fit for the different
$B-V$ ranges. In Fig. \ref{cyc_per_short_tau_c_dr_vs_rosby}, these regressions are shown 
as solid and coloured lines, and the respective fit parameters are listed in
Table \ref{results_final_short_cycle}.

We note that with this approach, HD~22049 is well fitted in the trend 
of the $B-V$ range $0.7<B-V\leq1.1$ and the fit results do not depend on whether or
not  HD~22049 is included or not. Based on these results, we modelled the observed
cycle periods using the expression:
\begin{eqnarray} \label{final_cycle_relation_short}
  P^{S}_{\rm{cyc}} & = & \frac{10^{a^{S}_{i}}}{d_{cz}} \tau_{c} Ro^{n^{S}_{i}} = 10^{a^{S}_{i}} v^{-1}_{dcz} Ro^{n^{S}_{i}},
\end{eqnarray}
where $10^{a^{S}_{i}}$ is a constant and $n^{S}_{i}$ is the exponent of the
Rossby number, and the index, $i,$ labels the different $B-V$ ranges. 
  
\begin{figure}  
\centering  
\includegraphics[scale=0.5]{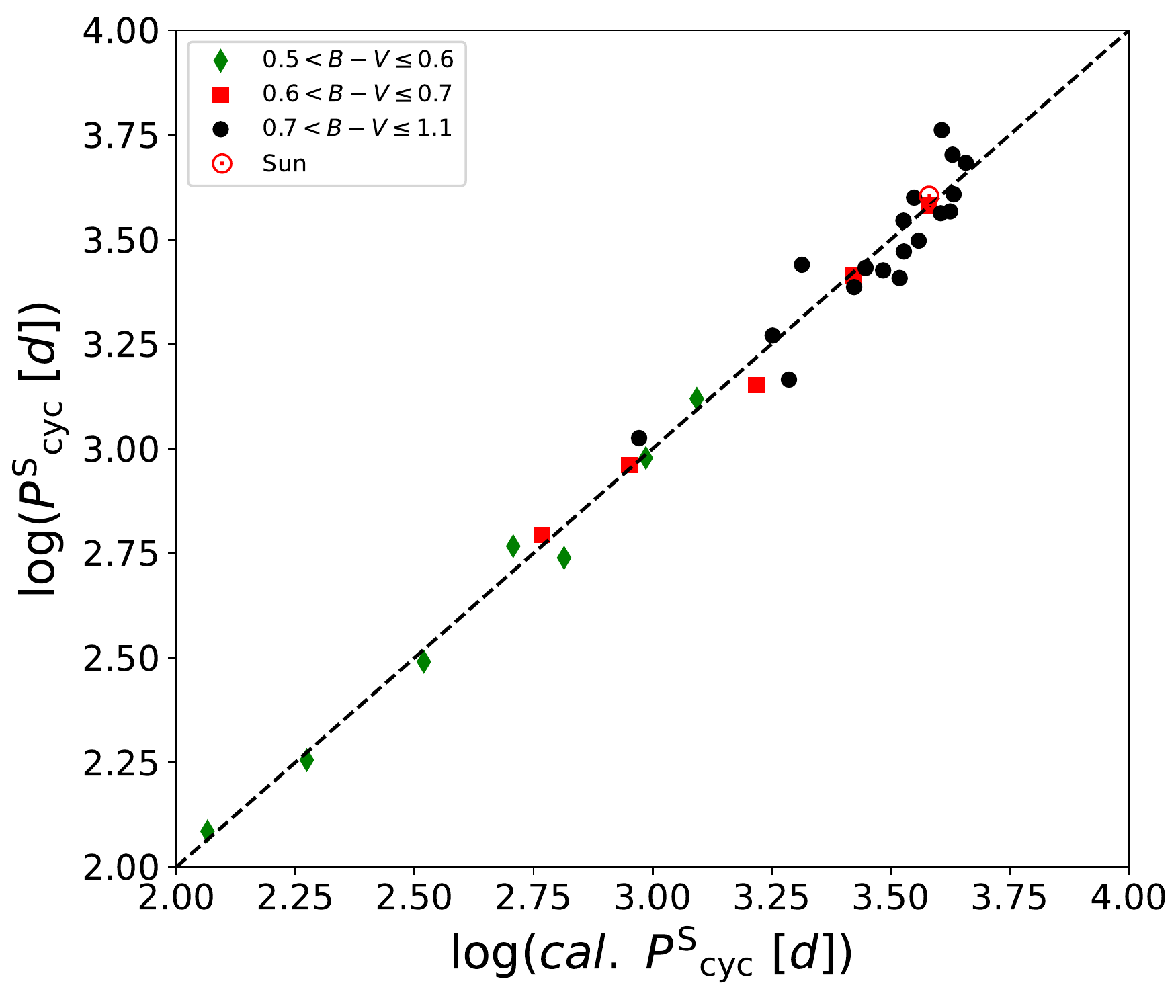}  
\caption{Deviations of the observed stellar cycles from the relation of
  Eq. \ref{final_cycle_relation_short}, the $B-V$ ranges are colour coded. The dashed line depicts the identity. }  
\label{compare_lin-ro_dr_tau}  
\end{figure}

\begin{table}[!t]  
\caption{Estimated parameters for Eq. \ref{final_cycle_relation_short_A} }  
\label{results_final_short_cycle}  
\begin{center}  
\begin{small}
\setlength{\tabcolsep}{5pt}
\begin{tabular}{l c c c c}  
\hline  
\hline  
\noalign{\smallskip}  
B-V range  &  No. of & $a^{S}$ & $n^{S}$ & $\sigma$ of \\
            & data &         &          & $\Delta$P$_{\rm{cyc}}^{S}/$P$_{\rm{cyc}}^{S}$ \\
\hline  
\noalign{\smallskip}  
$0.5<B-V\leq0.6$ & 7 & 1.53$\pm$0.05 & 1.93$\pm$0.11 & 0.10 \\
$0.6<B-V\leq0.7$ & 6 & 1.80$\pm$0.04 & 2.44$\pm$0.12 & 0.08 \\
$0.7<B-V\leq1.1$ & 18 & 1.48$\pm$0.04 & 1.09$\pm$0.12 & 0.17 \\
\hline  
%\tableline  
\end{tabular}
\addtolength{\tabcolsep}{3pt}
\end{small}  
\end{center}  
\end{table}  

In Fig.~\ref{compare_lin-ro_dr_tau}, the observed cycle periods are directly compared
with the calculated ones from the  derived best-fit relation
in Eq. \ref{final_cycle_relation_short} with the parameters listed in
Table~\ref{results_final_short_cycle}. These values are listed in Table \ref{tab1}.
In Fig.~\ref{compare_lin-ro_dr_tau}, the three $B-V$ ranges are
labelled with the same colour-coding as in Fig.~\ref{log_cyc_vs_log_ro}.
Figure~\ref{compare_lin-ro_dr_tau} demonstrates that the calculated cycle
periods are very well distributed around the identity line (plotted as a 
dashed line in Fig.~\ref{compare_lin-ro_dr_tau}). 

\begin{figure}  
\centering  
\includegraphics[scale=0.5]{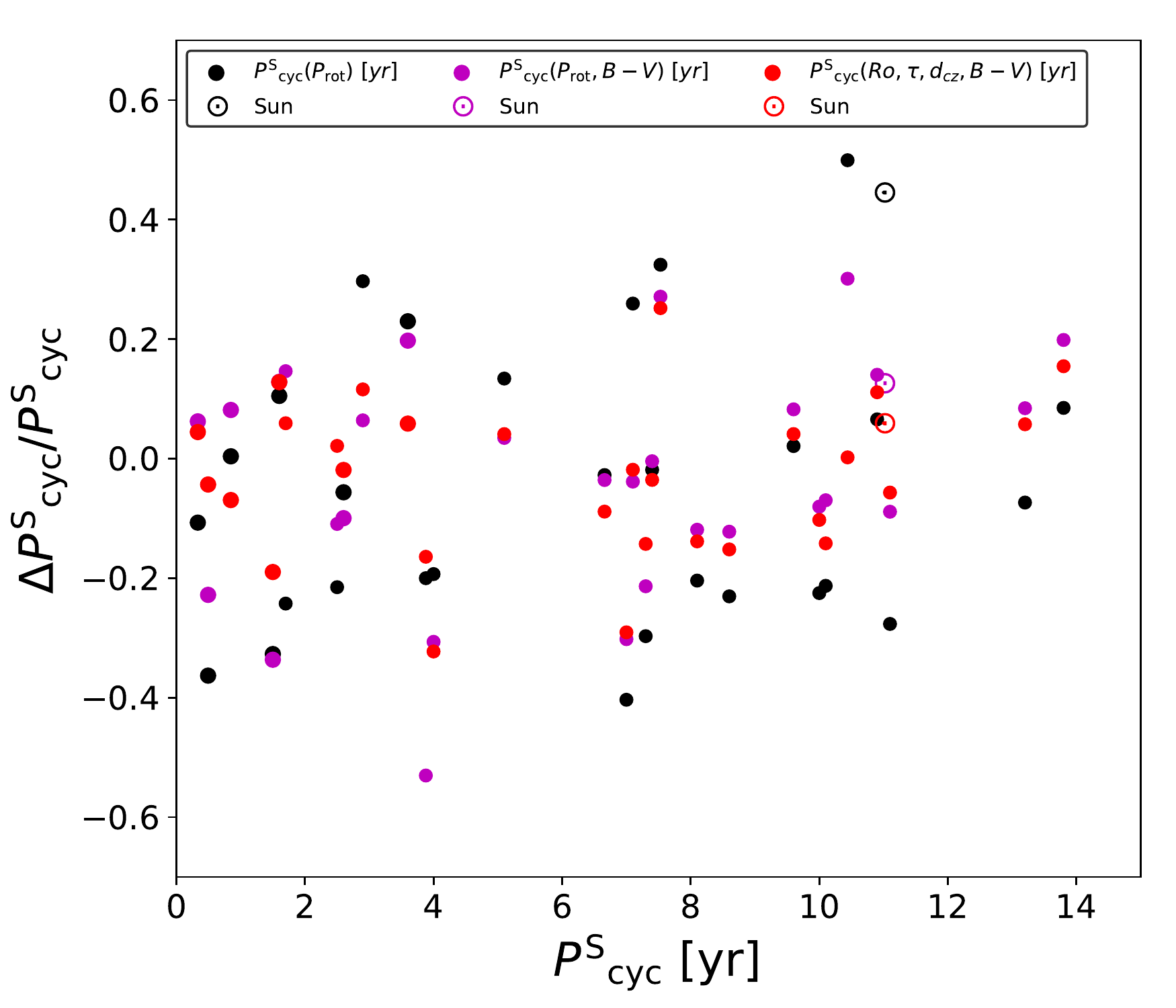}  
\caption{Relative deviations versus cycle on the inactive branch.
  Black data points show $\Delta P^{\rm{S}}_{\rm{cyc}}/P^{\rm{S}}_{\rm{cyc}}$
  for the relation given by Eq. \ref{pcyc-prot}, magenta data points the
  $\Delta P^{\rm{S}}_{\rm{cyc}}/P^{\rm{S}}_{\rm{cyc}}$ for the relation given by 
  Eq. \ref{lin_log_pcyc_log_ro_a} with parameter listed in Table~\ref{tab_trend_test} and 
  red data points $\Delta P^{\rm{S}}_{\rm{cyc}}/P^{\rm{S}}_{\rm{cyc}}$ for the relation given by 
  Eq. \ref{final_cycle_relation_short}.}
\label{compare_rel_dev}  
\end{figure}

Next, we determined the relative deviations
$\Delta P^{\rm{S}}_{\rm{cyc}}/P^{\rm{S}}_{\rm{cyc}}$  and show these values
as red data points in Fig. \ref{compare_rel_dev}. Furthermore,
we computed the corresponding standard deviation of $\Delta P^{\rm{S}}_{\rm{cyc}}/P^{\rm{S}}_{\rm{cyc}}$
to quantify the deviation between the observed and calculated cycle periods.
The standard deviations in the three $B-V$ range are listed in
Table \ref{results_final_short_cycle}.
For all stars in this study, a value of 0.14 was obtained, namely,
an average deviation of 14$\%$. This scatter is similar to the
one obtained around the relation given by Eq. \ref{lin_log_pcyc_log_ro},
implying no improvement by the inclusion of these
additional parameters. However, this may simply be a consequence of the small
data set available for this study.
Finally, we again calculated the solar activity cycle period by the
final empirical relation, which results in a calculated solar cycle
period of 10.3$^{1.1}_{1.0}$ yr. This value is fully consistent with the well-known 11-year Schwabe cycle.

In the following, we provide a brief comparison to the simple
cycle-rotation relation presented in Sects. \ref{cyc_per} and \ref{cyc_per_bv_ranges}.
Using only stars in the $B-V$ range of $0.5<B-V\leq1.1,$
we plot the relative deviation $\Delta P^{\rm{S}}_{\rm{cyc}}/P^{\rm{S}}_{\rm{cyc}}$
for the three relations in Fig.~\ref{compare_rel_dev}, where the black
data points show $\Delta P^{\rm{S}}_{\rm{cyc}}/P^{\rm{S}}_{\rm{cyc}}$
for the relation given by Eq. \ref{pcyc-prot}, magenta data points the
$\Delta P^{\rm{S}}_{\rm{cyc}}/P^{\rm{S}}_{\rm{cyc}}$ for the relation given by 
Eq. \ref{lin_log_pcyc_log_ro_a} with the parameters listed in Table~\ref{tab_trend_test}, and 
red data points $\Delta P^{\rm{S}}_{\rm{cyc}}/P^{\rm{S}}_{\rm{cyc}}$ for the relation given by 
Eq. \ref{final_cycle_relation_short}. It is obvious 
that $\Delta P^{\rm{S}}_{\rm{cyc}}/P^{\rm{S}}_{\rm{cyc}}$ obtained with
Eq. \ref{pcyc-prot} shows a larger scatter compared to the other. Also, we can see that $\Delta P^{\rm{S}}_{\rm{cyc}}/P^{\rm{S}}_{\rm{cyc}}$ obtained with 
Eq.~\ref{final_cycle_relation_short} show the smallest scatter.
We then compared the standard deviation of the
$\Delta P^{\rm{S}}_{\rm{cyc}}/P^{\rm{S}}_{\rm{cyc}}$ sets, 0.24 for
the relation Eq. \ref{pcyc-prot}, then 0.19 for Eq. \ref{lin_log_pcyc_log_ro_a}
and parameter listed in Table~\ref{tab_trend_test}, and 0.14 
for the relation in this section. These values are confirmed
the visual observation from Fig.~\ref{compare_rel_dev}. 

As a further test, the calculated cycle lengths for the solar cycle
can be considered:  with the simple linear relation Eq. \ref{pcyc-prot},
the relative  deviation of the Sun is $\approx$ 45$\%$. This stands in contrast
to the relative deviation of $\approx$ 13$\%$ obtained
with Eq. \ref{lin_log_pcyc_log_ro_a} and the parameters listed in Table~\ref{tab_trend_test}
as well as a relative deviation of $\approx$ 6$\%$ obtained
using the relation derived in this section. 

Finally, we tested the statistical significance of the variance reduction 
between the models. First, we compared the variance between 
the relation given by Eq. \ref{pcyc-prot} and the relation derived in this section via
an overall F-test, obtaining $F = 15.8$. Based on this F-value, we  
obtained a formal p-value of 4.6$\cdot10^{-10}$, which shows a clearly statistically
significant reduction in the residual deviations between these models.
Next, we compared the variance between the relation given by 
Eq. \ref{lin_log_pcyc_log_ro_a}, with the given parameter
from Table~\ref{tab_trend_test}, and the relation derived in this section. 
Both relations have the same numbers of degrees of freedom, so that we can use 
a F-test where the numbers of degrees of freedom are not taken into account.
We obtained an F-value of 1.91 and a corresponding formal p-value of 0.06. This p-value
shows a statistically significant reduction of 94$\%$ between these models.

\subsubsection{Summary and conclusions for our final $P^{S}_{\rm{cyc}}-Ro$ relation for the short-cycle branch}
In this section, we explain how we derived a cycle-Rossby number relation with and without
the $B-V$ splitting into three main ranges. In both relations, the
convective turnover time and the relative depth of the convection zone were considered.
We found that the relation without $B-V$ splitting does not describe the
cycle-rotation connection any better than the simple linear relation with the rotation
period does  (see Eq. \ref{pcyc-prot}). However, the relation derived
in Sect. \ref{final_cyc_ro_rel_sb}, which includes the splitting into
three $B-V$ ranges, shows the smallest deviation between the observed and calculated
cycle period compared to the other derived relations in this work. 
Therefore, we conclude that this new, final relation presented
in Sect. \ref{final_cyc_ro_rel_sb}, based on the Rossby number 
parameter, describes the cycle-rotation relation much better than a
simple linear relation with rotation period (see Eq. \ref{pcyc-prot}) and better than
the linear relation with rotation period including the $B-V$
splitting (see Eq. \ref{lin_log_pcyc_log_ro_a} and the parameters in Table~\ref{tab_trend_test}).

Furthermore, we conclude that the $B-V$ splitting into three ranges
is important for describing the cycle-rotation relation. This $B-V$
splitting into three main $B-V$ ranges is not removed with the
introduction of convective turnover time and the relative depth of the convection zone
or by the ratio of these two parameters. A comparison of the slopes given in
Tables \ref{tab2} and \ref{results_final_short_cycle} shows that the slopes
do not change significant significantly, supporting this conclusion.

\subsection{Cycles on the long-period branch versus Rossby number}
\label{norm_cyc2_ro}

\begin{figure}  
\centering  
\includegraphics[scale=0.45]{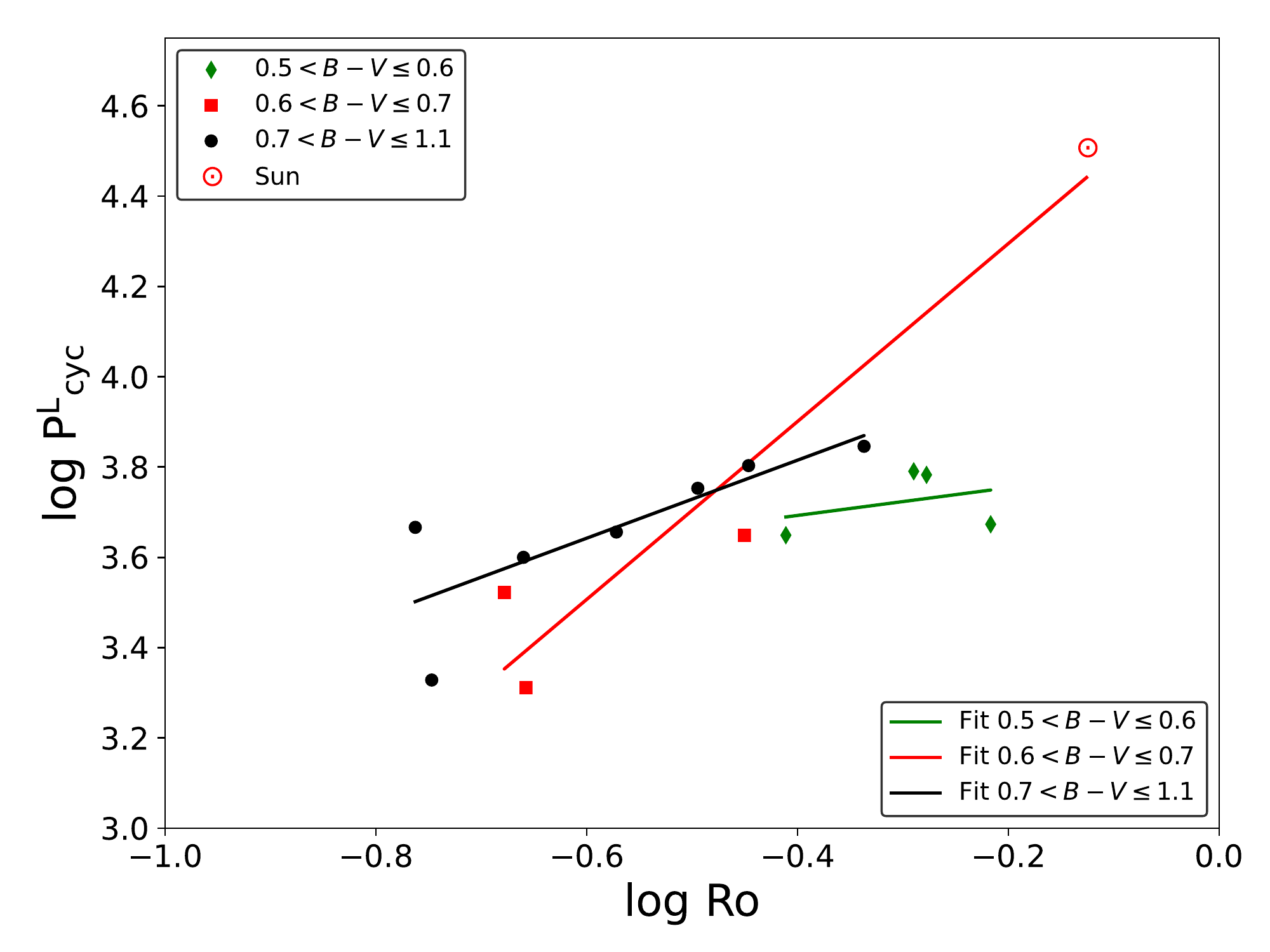}  
\caption{Logarithm of cycles on the long-cycle branch versus the logarithm
  of the Rossby numbers; different colours and symbols indicate
  data in different $B-V$ ranges. Solid coloured lines 
  show the different relations obtained for the fitted $B-V$ ranges.}  
\label{log_cyc2_vs_log_Ro}  
\end{figure}

In our sample, there are only 15 stars located on the long-cycle branch and
in the relevant $B-V$ range of $0.5<B-V<1.1,$ the available sample
is very small. Here, too, we split the data into the different
$B-V$ ranges and plot the logarithm of the cycle lengths on
the long-cycle (active) branch versus the logarithm of the Rossby numbers (see Fig.~\ref{log_cyc2_vs_log_Ro}).
A visual inspection of Fig.~\ref{log_cyc2_vs_log_Ro}
suggests some evidence for differences with $B-V$. 

To test and quantify any such $B-V$-specific relations, we focussed
again on three main $B-V$ ranges and performed a linear
least-squares fit (cf., Eq.\ref{lin_log_pcyc_log_ro});
the results of the fits are shown in Table~\ref{tab5}, including the
number of data points in the corresponding $B-V$ range.

\begin{table}[!t]  
\caption{Results of the $B-V$-specific relations for the cycles on the
  long-cycle branch}  
\label{tab5}  
\begin{center}  
\begin{small}  
\begin{tabular}{lccc}  
\hline  
\hline  
\noalign{\smallskip}  
$B-V$ range & No. of & A & n  \\  
& data & & \\
\hline  
\noalign{\smallskip}
$0.5<B-V\leq0.6$  & 4 & 3.82$\pm$0.18 & 0.31$\pm$0.60 \\
$0.6<B-V\leq0.7$  & 4 & 4.69$\pm$0.19 & 1.97$\pm$0.40 \\
$0.7<B-V\leq1.1$  & 7 & 4.16$\pm$0.17 & 0.86$\pm$0.29 \\
\hline  
%\tableline  
\end{tabular}
\end{small}  
\end{center}  
\end{table}

Given the scatter in the data on the long-cycle branch, as well as the paucity
of data points and the relatively small span of covered Rossby numbers,
we consider those relations relatively uncertain; this also applies
to the $B-V$ range of $0.5<B-V\leq 0.6$, where the slope is consistent with
zero. Furthermore, the relation for the $B-V$ range of $0.6<B-V\leq 0.7$
strongly depends on the adopted cycle value of the solar Gleissberg cycle. Therefore, we cannot obtain a meaningful  trend in  
this $B-V$ range based on stellar data alone.

Next, we tested whether the cycle periods located on the long-cycle branch
can be described in the same way as above for the short-cycle branch (see Sect. \ref{final_cyc_ro_rel_sb}). In this case, the corresponding relation is:
\begin{eqnarray} \label{final_cycle_relation_long}
  P^{L}_{\rm{cyc}} & = & \frac{10^{a^{L}_{i}}}{d_{cz}} \tau_{c} Ro^{n^{L}_{i}} = 10^{a^{L}_{i}} v^{-1}_{dcz} Ro^{n^{L}_{i}},
\end{eqnarray}
where (again) $a^{L}_{i}$ is a constant, $n^{L}_{i}$ is the exponent of the Rossby
number, and the index, $i,$ labels the different $B-V$ ranges. As described in Sect. \ref{final_cyc_ro_rel_sb},
we used the relations in Eqs. \ref{rel_v_bv1} and \ref{rel_v_bv2} to estimate the ratio of
the convective turnover time and relative depth of the convection.
Since the number of cycles on the long-cycle period branch is very limited, it is not possible
to fit the data the same way as in Sect. \ref{constant_a_est} to obtain robust parameters
for $a^{L}_{i}$ and $n^{L}_{i}$.

Instead of estimating the parameter $n^{L}_{i}$ via a fit, we used the parameters
$n^{S}_{i}$ of the short-cycle branch, based on the fact that the slopes $n$ in the
Eq. \ref{lin_log_pcyc_log_ro} are the same as those for the cycles
on the short- and long-cycle branch (see Tables \ref{tab2}, \ref{results_final_short_cycle},
and \ref{tab5}) within the error margins, implying that an individual fit will not
produce any significantly different values.

With the given $n^{S}_{i}$ values in Table \ref{results_final_short_cycle},
Eq. \ref{final_cycle_relation_long} can be transformed to $a^{L}_{i}$ and
by averaging the individual values of $a^{L}_{i}$ in the
corresponding $B-V$ ranges, the constants, $a^{L}_{i}$,  can be obtained for the
three used $B-V$ ranges. These results are listed in Table \ref{results_final_long_cycle}.

\begin{table}[!t]  
\caption{Estimated parameters for Eq. \ref{final_cycle_relation_long} }  
\label{results_final_long_cycle}  
\begin{center}  
\begin{small}
\setlength{\tabcolsep}{5pt}
\begin{tabular}{l c c  c c}  
\hline  
\hline  
\noalign{\smallskip}  
B-V range  & No. of & $a^{L}$ & $n^{L}$ & $\sigma$ of \\
            & data points &       &          & $\Delta$P$_{\rm{cyc}}^{L}/$P$_{\rm{cyc}}^{L}$ \\
\hline  
\noalign{\smallskip}  
$0.5<B-V\leq0.6$  & 4 & 2.31$\pm$0.14 & 1.93$\pm$0.11 & 0.37\\
$0.6<B-V\leq0.7$  & 4 & 2.80$\pm$0.17 & 2.44$\pm$0.12 & 0.34\\
$0.7<B-V\leq1.1$  & 7 & 2.00$\pm$0.10 & 1.09$\pm$0.12 & 0.24 \\
\hline  
%\tableline  
\end{tabular}
\addtolength{\tabcolsep}{3pt}
\end{small}  
\end{center}  
\end{table}  

\begin{figure}  
\centering  
\includegraphics[scale=0.5]{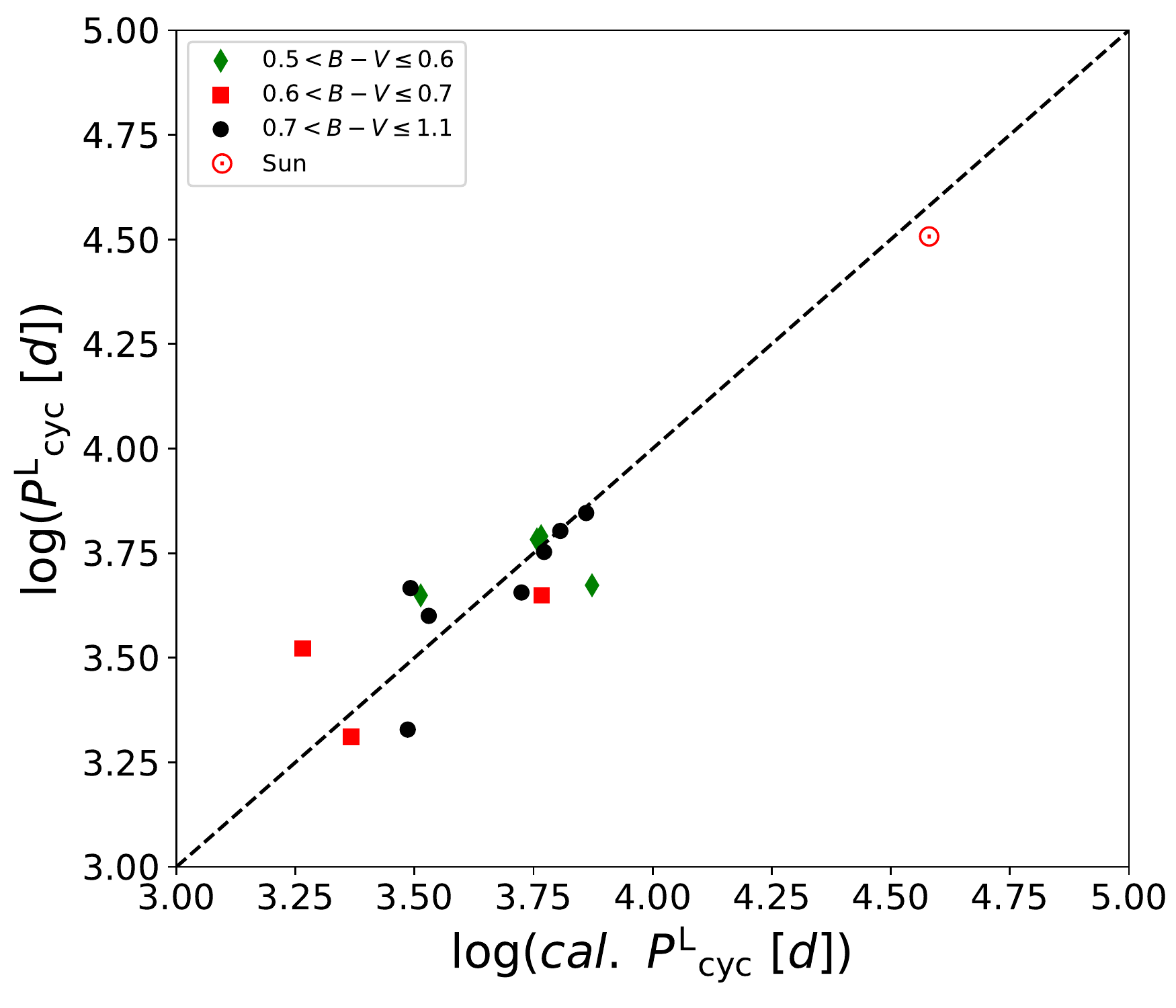}  
\caption{Comparison of the measured cycle periods on the long-cycle branch
  against the empirical relation given by Eq. \ref{final_cycle_relation_long}
  with the parameters listed in Table \ref{results_final_long_cycle},
  where the $B-V$ ranges are colour-coded. The dashed line depicts the
  identity relation. }  
\label{cyc_per_long_vs_rosby}  
\end{figure}

In Fig.~\ref{cyc_per_long_vs_rosby}, the measured cycle periods on the long-cycle branch are
compared with the calculated cycle periods
(as listed in Table \ref{tab1}) from the relation given by
Eq. \ref{final_cycle_relation_long}, with
the derived parameters listed in Table \ref{results_final_long_cycle}.
Inspection of these data shows they are well distributed around the
identity shown as dashed line.
Not unexpectedly, the scatter of these data is larger than for the data
on the short-cycle period branch: we compute 
a standard deviation of 0.28, a factor $\sim$2 larger
than the standard deviation found on the short-cycle period branch
with our final, best empirical relation.
Finally, we compare the Gleissberg cycle of the Sun with the expectation
given by Eq. \ref{final_cycle_relation_long} and obtain $104^{+50}_{-34}$ yr.
This value deviates by only 17$\%$ from the length of the Gleissberg
cycle of 88 years according to \citep{Ptitsyna2021Ge&Ae..61S..48P}, and lies within the
range of values suggested by different authors. Furthermore, 
we point out the large error range of our calculated
Gleissberg cycle of the Sun, which is a reflection of the large uncertainty
of the cycle relation in this cycle branch due to the small number of data on this
cycle branch.

\section{Comparison with previous studies}
\label{compare_with_pre_studies}

In this section, we provide a brief comparison of our results for the short-cycle branch
with those obtained in previous studies. As pointed out before, the main difference in 
our approach is that we are looking for a relation between Rossby number and
cycle periods, including possible additional parameters; our approach is, in some ways,
similar to that of \citet{Noyes1984ApJ...287..769N}, who used solar-like cycles 
obtained from early Mount Wilson data \citep{wilson1978} and related
the inverse cycle period and the inverse rotation period, thus finding 
an additional $B-V$-dependence of the emerging empirical relation, which they could
interpret in terms of the convective turnover time.

In this context, we has to keep in mind that the stellar sample used 
by \citet{Noyes1984ApJ...287..769N} contained only 13 stars (including
the Sun) and for 6 stars, they  used only derived (not observed) rotation periods.
Nevertheless, when we study the equivalent relation with our far larger stellar data
(see Eq. \ref{pcyc-prot}),
we derived  $P_{\rm{cyc}} \propto P_{\rm{rot}}^{1.324\pm0.067}$,
with an exponent that is very comparable to that already obtained by \citet{Noyes1984ApJ...287..769N}.

Another approach to studying the activity cycle versus rotation period is given
by $\omega_{\rm{cyc}}/\Omega = P_{\rm{rot}}/P_{\rm{cyc}}$, as
used by \citet{Brandenburg1998ApJ}. That work already showed that the logarithm
of $\omega_{\rm{cyc}}/\Omega$ is correlated with the logarithm of the inverse Rossby 
number (thus implying an exponential relation) or, rather, the activity index, $\log R_{\rm{HK}}^{\prime}$.
In these relations, the two cycle branches had become visible.
 
With a larger and updated sample, \citet{Brandenburg2017} used the expression:
\begin{eqnarray} \label{brandenburg1}
\frac{\omega_{\rm{cyc}}}{\Omega} & = & b_{i}\langle R_{\rm{HK}}^{\prime}\rangle^{\nu_{i}}
,\end{eqnarray}
to estimate the activity cycles, where the index, i, distinguishes the two branches:
A for active (long) and I inactive (short) cycles.
The factor $b_{i}$ is a fit parameter not specified by 
\citet{Brandenburg2017}. The authors justified this with the argument that the
value of $\omega_{\rm{cyc}}/\Omega$ does physically not exist for 
$\log R_{\rm{HK}}^{\prime} = 0$. Instead of $b_{i}$ the value $\log \tilde{b}_{i}$ is
specified, where $\tilde{b}_{i}$ is $\omega_{\rm{cyc}}/\Omega$ at the Vaughan-Preston gap
with $\log R_{\rm{HK}}^{\prime} = -4.75$.
\begin{figure}  
\centering  
\includegraphics[scale=0.43]{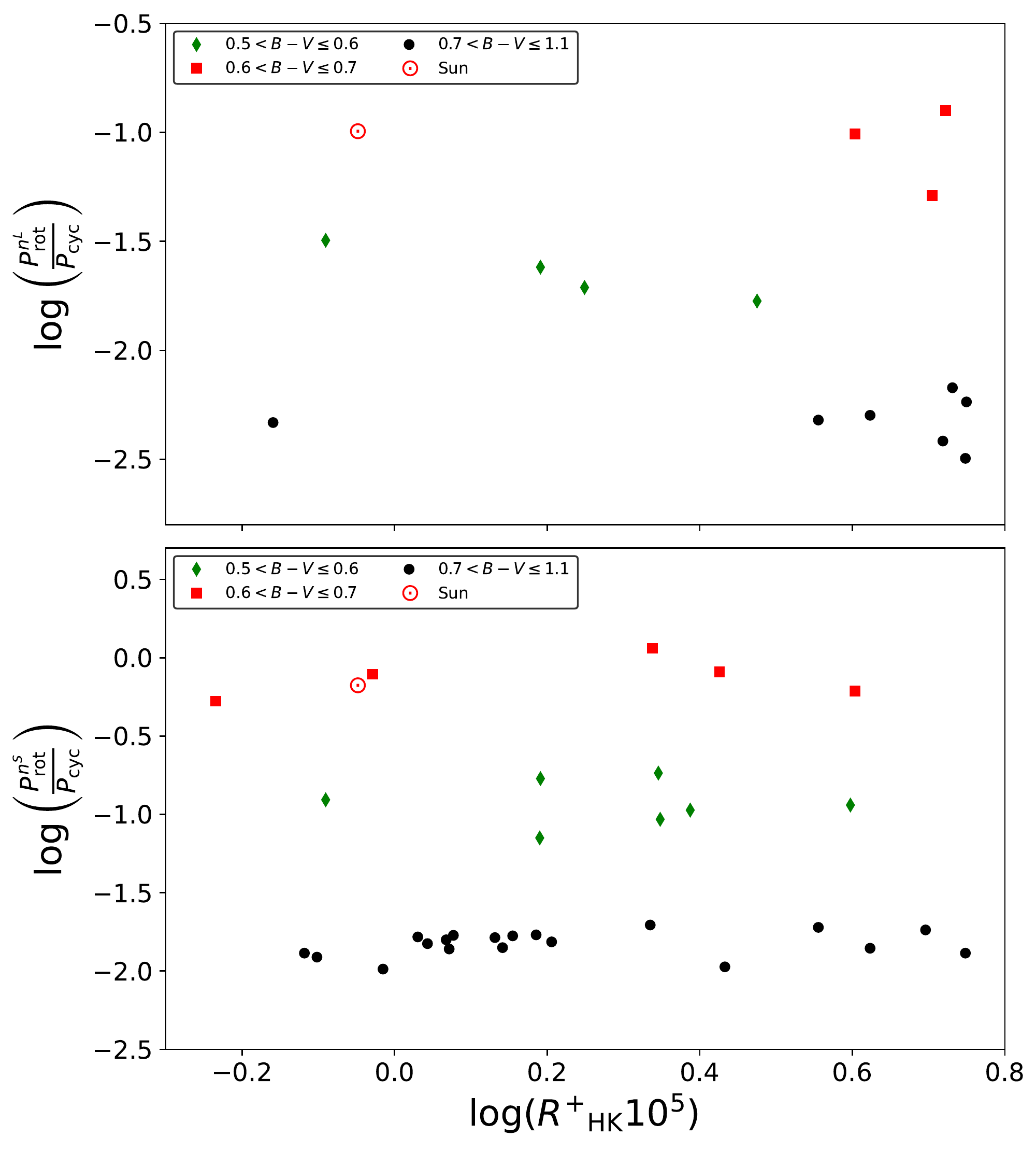}  
\caption{Logarithmic ratio between rotation period and activity
    cycles over an activity scale given
    by $\log R_{\rm{HK}}^{+}10^{5}$ and the different colours and symbols
    distinguish different $B-V$ ranges. Upper panel: Results for the cycles on the
    long-cycle branch. Lower panel: Results for the cycles on the
    short-cycle branch.}
\label{log_prot_exp_n-pcyc_rhk}  
\end{figure}

The correlation between the $\omega_{\rm{cyc}}/\Omega,$ or rather $P_{\rm{rot}}/P_{\rm{cyc}}$
and the Rossby number or the activity index, $R_{\rm{HK}}^{\prime}$, is equivalent to the empirical 
relation of the form  $P_{\rm{cyc}} \propto P_{\rm{rot}}^{\rm{n}}$. 
On this basis, the ratio of the rotation period and cycle period is
\begin{eqnarray} \label{frac_prot_pcyc_com}
\frac{\omega_{\rm{cyc}}}{\Omega} & = & \frac{P_{\rm{rot}}}{P_{\rm{cyc}}} \propto \frac{1}{P_{\rm{rot}}^{n-1}}.
\end{eqnarray}

Therefore, we assume that the activity index $R_{\rm{HK}}^{\prime}$ in the proportionality 
$\omega_{\rm{cyc}}/\Omega = P_{\rm{rot}}/P_{\rm{cyc}} \propto R_{\rm{HK}}^{\prime}$ (see Eq. \ref{brandenburg1})
is caused by a remaining proportionality of the activity-rotation relation. Thus, we do not
expect any dependence between the ratio, $P_{\rm{rot}}^{n}/P_{\rm{cyc}}$, and activity index
$R_{\rm{HK}}^{+}$, which we use instead of $R_{\rm{HK}}^{\prime}$ and which is the pure \ion{Ca}{ii}~H\&K flux excess.
Consequently, the following applies:\ \begin{eqnarray} \label{frac_prot_pcyc_com_a}
\frac{P_{\rm{rot}}^{n^{S}_{i}}}{P_{\rm{cyc}}}(R_{\rm{HK}}^{+}) & \propto & constant, 
\end{eqnarray}
and
\begin{eqnarray} \label{frac_prot_pcyc_com_b}
\frac{P_{\rm{rot}}^{n^{L}_{i}}}{P_{\rm{cyc}}}(R_{\rm{HK}}^{+}) & \propto & constant, 
\end{eqnarray}
where n is the exponent of the empirical proportionality, the index $S$ and $L$ 
labelling the two cycle branches, and the index, $i,$  represents different
$B-V$ ranges. 

To test this assumption, we computed the period ratio (cycle over rotation) as in 
Eqs. \ref{frac_prot_pcyc_com_a} and \ref{frac_prot_pcyc_com_b},
for both branches, and we used the results obtained in Sects. \ref{norm_cyc_ro} and \ref{norm_cyc2_ro} (see
Tables \ref{results_final_short_cycle} and \ref{results_final_long_cycle}). 
 Figure~\ref{log_prot_exp_n-pcyc_rhk} visualises these results over an 
activity scale in double-logarithmic figures, colour-coding the $B-V$ ranges. 
Indeed, here the activity seems to make no difference in any single $B-V$ range, 
which gives empirical support to the above relation. 
Yet there are evident differences between different $B-V$ ranges.  

\begin{figure}  
\centering  
\includegraphics[scale=0.5]{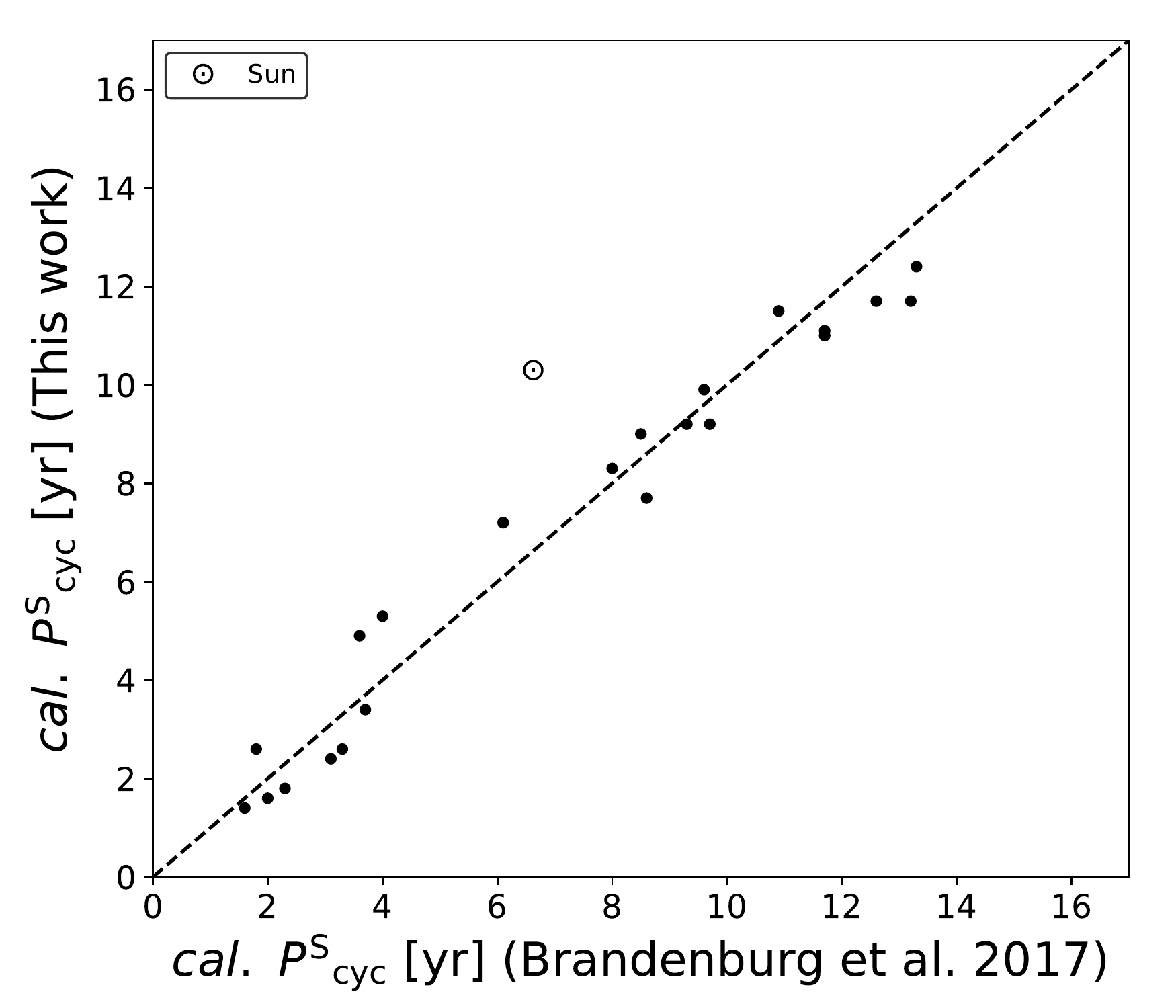}  
\caption{Calculated $P^{S}_{\rm{cyc}}$ obtained from the empirical relation
  found in this study over the respective values from \citet{Brandenburg2017};
  the dashed line depicts the identity.}  
\label{compare_our_with_brandenburg}  
\centering  
\includegraphics[scale=0.5]{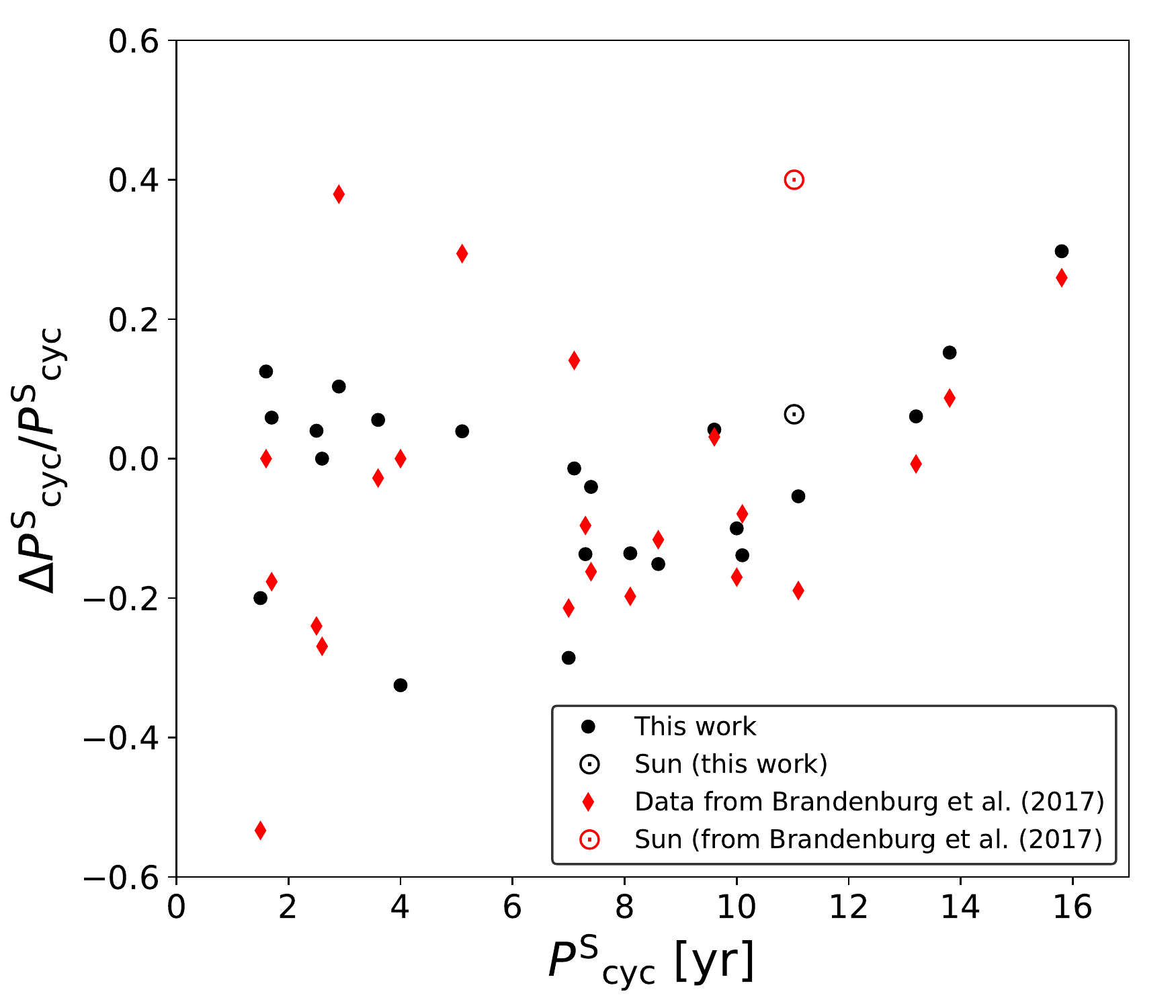}  
\caption{Relative deviation versus cycle on the short-cycle branch for 23 stars contained
  in both samples; black data points are obtained with our calculated cycle periods,
   red diamonds from \citet{Brandenburg2017}.}  
\label{relative_residuals_our-brandenburg}  
\end{figure}

Finally, in Fig.~\ref{compare_our_with_brandenburg} we compare the
cycle periods from our final empirical relation with a Rossby number
with those periods from \citet{Brandenburg2017} for 23 stars located on the short-cycle
branch, which are contained in both samples, except HD~103095; this is because the
calculated cycle listed in Table \ref{tab1} is computed with Eq. \ref{lin_log_pcyc_log_ro} and
the corresponding parameter given in Table \ref{tab2}. This comparison shows a good agreement between both
sets of results, except for the Sun.

Comparing the expectations of the 11-year solar cycle given by the empirical relation
of \citet{Brandenburg2017} and the one presented here, we find that the estimated
solar cycle of 6.6 years by \citet{Brandenburg2017} clearly differs
from the observed eleven year solar cycle, while our estimated solar cycle period of 10.3 years
is consistent with the 11-year solar cycle.
This might be related to our use of smaller $B-V$ ranges, as well as the use of the
Rossby number instead of the rotation period.
Also, we show the relative deviations of the compared cycles in Fig. \ref{relative_residuals_our-brandenburg},
and there we can observe a larger scatter of the values obtained by the cycle-activity relation
of \citet{Brandenburg2017} than in our final cycle-rotation relation. To quantify this, we computed the corresponding
standard deviations of the relative deviations and obtained a standard deviation of 0.22 for the values of the
cycle-activity relation of \citet{Brandenburg2017} and a standard deviation of 0.14 for our final cycle-rotation relation.

Furthermore, we performed again a overall F-test (like in Sect. \ref{sum_diff_approaches}) 
to see whether both empirical relations are actually statistically different in a
significant way. For the F value, we obtained $F=3.79$, yielding a 
confidence level of $99.6\%$. Hence, we can assume that the our final cycle-rotation
relation is indeed statistically significant in its difference from the 
one given by \citet{Brandenburg2017},  and the reduction of the residual differences 
of the individual data points is real.

\section{Depth of dynamo action}
\label{depth_est}

At least some stars in our sample do show two activity cycles: one on the short-cycle branch
and one on the long-cycle branch, thus raising the question whether these different activity cycles may be
produced by dynamos operating in different depths of the convective zone. 
A common assumption made by the mean field $\alpha-\Omega$ dynamo theory is that the dynamo action
takes place at the bottom of the convection zone. In that case, the relative length scale of the convection
could also be seen as representative for the relative depth of dynamo action.

Here, we follow the approach to use the relative depth of the convection zone ($d_{cz}$) instead of
the relative length scale of the turbulence ($l/R_{\star}$) as an additional parameter for our cycle relations
(see Eqs. \ref{final_cycle_relation_short} and \ref{final_cycle_relation_long}).
Furthermore, we assume that the constant $a$ in Eqs. \ref{final_cycle_relation_short} \& \ref{final_cycle_relation_long}
is a scaling factor of the cycle depth.
Consequently, we can write
\begin{eqnarray} 
\label{relative_depth_assumtion}
 \left(\frac{l^{j}_{i}}{R_{\star}}\right) & = & \frac{d_{cz}}{10^{a^{j}_{i}}},
\end{eqnarray}
where index $i$ distinguishes the different $B-V$ ranges
and index $j$ is the cycle branch. With this assumption, it is possible to assess
the depths  where the short and long cycles are produced. 
Calculating the ratio between the cycle periods located on the
long-cycle branch and the short-cycle branch (see Eqs. \ref{final_cycle_relation_short} and \ref{final_cycle_relation_long}) 
and using our assumption on the constant a (see Eq. \ref{relative_depth_assumtion}), we express the
ratio between the periods on the long- and short-cycle branches as:
\begin{eqnarray} \label{const_cycle_2}
  \frac{P^{L}_{\rm{cyc}}}{P^{S}_{\rm{cyc}}} & = & \frac{10^{a^{L}_{i}} d_{cz} \tau_{c} Ro^{n^{L}_{i}}}{10^{a^{S}_{i}} d_{cz} \tau_{c} Ro^{n^{S}_{i}}} = \frac{l^{S}_{i} \tau_{c} Ro^{n^{L}_{i}}}{l^{L}_{i} \tau_{c} Ro^{n^{S}_{i}}},
\end{eqnarray}
where index $L$ denotes the long-cycle branch and index $S$ is the short-cycle branch,
while index $i$ again labels the different $B-V$ ranges. For the
cycle-rotation relation on the long-period branch (see Sect  \ref{norm_cyc2_ro}), 
we assume that the exponent $n$ is the same in both branches, hence $n^{L}_{i} =n^{S}_{i}$.
That is supported by our empirical results (see Tables \ref{tab2} and
\ref{tab5}), where the values of $n$ are equal (within the uncertainties) in 
corresponding $B-V$ ranges.

With this assumption, the terms of $\tau_{c}Ro^{n^{L}_{i}}$ and $\tau_{c}Ro^{n^{S}_{i}}$ cancel
each other out, so that the ratio
of the cycle periods (see Eq. \ref{const_cycle_2}) is reduced to:
\begin{eqnarray} \label{const_cycle_3}
  \frac{P^{L}_{\rm{cyc}}}{P^{S}_{\rm{cyc}}} & = & \frac{ l^{S}_{i} }{ l^{L}_{i}}.
\end{eqnarray}
Equation~\ref{const_cycle_3} shows that the cycle ratio is only proportional to the 
ratio of the length scale of turbulence $(l)$.
Since cycle periods on the long-period branch are -- by definition -- longer than
on the short-cycle branch,
the length scale of turbulence $(l)$ must satisfy 
\begin{eqnarray} \label{const_cycle_5}
  l^{S}_{i} & > & l^{L}_{i}.
\end{eqnarray}
We therefore conclude that the cycles on the short-cycle period branch are created in layers of 
the convective zone that are deeper than those where the cycles on the long-cycle branch are produced.

\section{Summary and conclusions}
\label{discussion}

In this paper, we revisit the activity cycle-rotation connection for cool
main sequence stars with the aim of deriving a universal relation by
considering the impact of further stellar parameters, which would
be fully consistent with the solar cycle and rotation periods. In particular,
we consider the Rossby number as the natural parameter in any such relation,
which practically means a `normalisation' of any stellar rotation period
by the respective convective turnover time.

Studying the empirical relation between cycle and rotation periods with
different approaches, the statistically best empirical description
turns out not to be a linear relation -- but that of a power law (i.e. linear in
the double-logarithmic representation) and an additional $B-V$ dependence
must be considered. As suggested by earlier studies, we actually find
a bifurcation into a longer and a shorter cycle period branch. For the
latter, in this relation, we obtained an average scatter
of $\approx$ 20$\%$ among the sample stars.

This scatter can be decreased further by using the Rossby number instead
of the rotation period. By considering the convective turnover time
as well as the relative depth of the convection zone, we find that
the relative deviation between the actual observed and nominal cycle
periods decreases to 14$\%$. A further confirmation comes from
the Sun, where the calculated cycle length
of $10.3^{1.1}_{1.0}$~yr as obtained with our cycle period versus Rossby number
relation is fully consistent with the well-known 11-year solar activity cycle.

With the same approach, we also derived a cycle period-Rossby number relation
for the long-period branch. Here, we obtained an average scatter of 28$\%$
in the relative deviations, taking the original stellar data against the
empirical relation. For the Sun, this branch suggests a (Gleissberg) cycle
length of $104^{+50}_{-34}$~yr. This value is on the larger side of the
ranges of the suggested Gleissberg cycle periods, which are typically around 90 years,
yet given the stellar scatter and its uncertainties, we consider
the agreement of our calculated long-period branch with the observed Gleissberg cycle length
as satisfactory.

In summary, the solar Schwabe and Gleissberg cycles are reproduced very well
in our cycle-Rossby number relation in contrast to the cycle versus rotation representation.
In previous studies, the position of the Schwabe cycle in the cycle versus rotation presentation
is explained with the assumption of another kind of dynamo, for instance, a transitional dynamo. However, in
our study, we have not found any evidence suggesting another kind of dynamo and our
approach allows for the interpretation of both the Schwabe and Gleissberg cycles with the same relations
that apply to our cool main sequence sample stars on the short- and long-cycle branches.
Furthermore, in our sample, we found no objects with a strong deviation in the cycle versus 
Rossby number relation, which would suggest other dynamos or processes to be at work.

In our study, we also tested whether the cycle period is dependent on the
strength of the activity level. For this purpose, we compared
the logarithm of the ratio $P_{\rm{rot}}^{n}/P^{S}_{\rm{cyc}}$ with the
logarithm of activity index $R_{\rm{HK}}^{+}10^{5}$.
By using a power law approach, $P_{\rm{rot}}^{n}$, we avoided a secondary
dependency introduced artificially by the rotation--activity relation.
Our results suggest that the ratio $P_{\rm{rot}}^{n}/P^{S}_{\rm{cyc}}$
shows no dependence on the activity index $R_{\rm{HK}}^{+}10^{5}$ and,
therefore, we conclude that the cycle periods are indeed independent
of the activity level, which means there is thus no additional factor in our
cycle-rotation or rather cycle-Rossby number
relation (in contrast to the aforementioned $B-V$ colour index).

Finally, we briefly discuss indicators for the depth of the respective
dynamo operation. In our empirical relation between cycle period and
Rossby number, we used the relative depth of the convection zone in place
of the relative length scale of the turbulence, which can also be
interpreted as a relative depth of the dynamo operation.

From our factor analysis, we found that this relative depth of the
convection zone enters with an exponent that is away from unity, which
suggests that the dynamo is not created on the bottom the convention
zone. However, it should be mentioned here that the derived
exponents depend strongly on the other parameters in our relations.
Nevertheless, it seems feasible to make an estimation of which cycle
type is created in  deeper layers. Those located on the short-period branch
are probably created by a dynamo operating in deeper layer than those
to be associated with cycles located on the long-period branch. 

Clearly, the main problem in the entire analysis of the relation between the
cycle length and rotation period is the limited number of well known active
cycles. Due to this limited number of activity cycles with well-known periods,
the derived empirical relations have significant uncertainties. At least
the one for stars of solar colour matches the 11-year solar cycle relatively
accurately. Still, to verify the solar cycle relation and the other
relations, it remains important to find more reliable cycle periods,
especially among the long-period cycles. This is a task which requires
long-term time series observations. The only way to obtain time series of 
30 years and more  is to combine observations
from different monitoring programs. Hence,  \ion{Ca}{II} H\&K monitoring programs
are required to study the cycle-rotation relation, as a larger
number of calibration stars exist as well as recipes to ensure consistency between
very different instrumentation, epochs, and index definitions. 

Finally, we like to comment on the denotation of the two cycle branches as
`inactive' and `active', introduced by earlier studies. What is evident is
only a bifurcation into short- and long-period branches, as we confirm
here for our empirical relations over the Rossby number. Furthermore, our
Sun has two activity cycles, where one is located on the
`inactive' (short-cycle) branch and one on the `active' (long-cycle)
branch. However, we could show that the ratio of $P_{\rm{rot}}^{n}$ and $P_{\rm{cyc}}$
is independent of the strength of activity, including cases where the
same star has one cycle on each of those two branches.
Hence, we conclude that the bifurcation into those different branches in
the diagram of cycle length over rotation period is actually not caused
by the activity level of a star, but arises from a difference in depths
of dynamo operation in the convection zone, where the respective cycles
are created.

In consequence, we think that the hitherto common denotation of the two
cycle branches is a misnomer.
In \citet{Brandenburg2017}, where the co-existing of two activity
cycles are investigated, the `inactive' branch was renamed as short-cycle
branch and the `active' branch as long-cycle branch. We argue that this
denotation of the two branches is better rooted in reality and thus we suggest that the terms introduced by \citet{Brandenburg2017}
for the two branches should be used in the future.

\begin{acknowledgements}
We thank an anonymous referee for very helpful comments.
This research has made valuable use of the VizieR catalogue access tool,   
CDS, Strasbourg, France. The original description of the VizieR service   
was published in A\&AS 143, 23. We grateful for travel support and other
expenses covered  by bilateral project funding of Conacyt and DFG under
grant number 287156, as well as by support from our universities.
\end{acknowledgements}

\bibliographystyle{aa}  
\bibliography{bibfile.bib}  

\end{document}